\documentclass[letterpaper,twocolumn,10pt]{article}
\usepackage{usenix2019_v3}

\usepackage{tikz}
\usepackage{amsmath}

\usepackage{filecontents}

\usepackage{times}

\usepackage{soul}
\usepackage{url}
\usepackage[utf8]{inputenc}
\usepackage[small]{caption}
\usepackage{graphicx}
\usepackage{amsmath,amssymb,amsfonts}
\usepackage{booktabs}
\usepackage{textcomp}
\usepackage{wrapfig}
\usepackage{subfigure}
\usepackage{comment}
\usepackage{xcolor}
\usepackage{tcolorbox}

\usepackage{tabularx}
\usepackage{lscape}
\usepackage{mathtools}
\usepackage{booktabs}
\usepackage{algpseudocode}
\usepackage{algorithm}

\DeclareMathOperator*{\argmin}{arg\,min}
\usepackage{authblk}
\begin{document}
%-------------------------------------------------------------------------------

%don't want date printed
\date{}

%\title{\Large \bf a-RNA: Adversarial Radio Noise Attack to Fool Radar-based Environment Perception Systems}
\title{\Large \bf Adversarial Attack on Radar-based Environment Perception Systems}

%for single author (just remove % characters)
%\author{
%{\rm Amira Guesmi, Ihsen Alouani}\\
%IEMN CNRS-UMR 8520, Université Polytechnique Hauts-de-France\\

\author[1,2]{Amira Guesmi}
\author[1,3]{Ihsen Alouani}
\affil[1]{IEMN CNRS-UMR 8520, Université Polytechnique Hauts-de-France}
\affil[2]{eBrain Laboratory, New York University Abu Dhabi, UAE}
\affil[3]{CSIT, Queen's University Belfast, UK}

\renewcommand\Authands{ and }
%\and
%{\rm Second Name}\\
%Second Institution
% copy the following lines to add more authors
% \and
% {\rm Name}\\
%Name Institution
%} % end author

\maketitle

%-------------------------------------------------------------------------------
\begin{abstract}

%Ultra-Wide Band (UWB) radars have been utilized in different environment perception applications such as intelligent transportation systems (ITS). 
%Obstacle detection and recognition is a safety critical task for intelligent transportation systems. 
Due to their robustness to degraded capturing conditions, radars are widely used for environment perception, which is a critical task in applications like autonomous vehicles. More specifically, Ultra-Wide Band (UWB) radars are particularly efficient for short range settings as they carry rich information on the environment. 
Recent UWB-based systems rely on Machine Learning (ML) to exploit the rich signature of these sensors. However, ML classifiers are susceptible to adversarial examples, which are created from raw data to fool the classifier such that it assigns the input to the wrong class. These attacks represent a serious threat to systems integrity, especially for safety-critical applications. Several adversarial attacks have been developed during the recent years targeting different application domains such as computer vision, speech recognition, healthcare, etc. While these works highlighted the vulnerability of ML systems to adversarial noise, few of the underlying attack scenarios are practical in real-life. %The difficulty comes mainly from  This work tackles a realistic case of injecting  first 

In this work, we present a new adversarial attack on UWB radars in which an adversary injects adversarial radio noise in the wireless channel to cause an obstacle recognition failure. First, based on signals collected in real-life environment, we show that conventional attacks fail to generate robust noise under realistic conditions. We propose a-RNA, i.e., Adversarial Radio Noise Attack to overcome these issues. Specifically, a-RNA generates an adversarial noise that is efficient without synchronization between the input signal and the noise. Moreover, a-RNA generated noise is, by-design, \textbf{robust} against pre-processing countermeasures such as filtering-based defenses. Moreover, in addition to the undetectability objective by limiting the noise magnitude budget, a-RNA is also efficient in the presence of sophisticated defenses in the spectral domain by introducing a frequecy budget. 
%In fact, the a-RNA noise is generated under randomized incidence time and by not only clipping in the noise magnitude budget, but also in the frequency range. % and averaged on random noise incidence time in the receiver

%The proposed setting is based on signals collected in real-life environment, and we show that even without synchronization between the input signal and the noise, the adversarial attack is still effective. RNA generates, by-design, an adversarial noise in the same frequency range of the victim signal to avoid detection and evade filetering-based defenses. %We show that a pre-processing defense can significantly limit the effectiveness of the attack, but we propose an adaptive attack that bypasses the pre-processing defense. 
We believe this work should alert about potentially critical implementations of adversarial attacks on radar systems that should be taken seriously.

\end{abstract}

\section{Introduction}\label{sec:intro}

Intelligent Transportation Systems (ITS) and robotics drove tremendous effort in both industry and research community that led to promising ML-driven environment perception and scene understanding solutions. One of the major challenges in building high-performance environment perception for ITS is designing robust obstacle detection/recognition systems. %Even with the recent advances in ML, environment perception still remains a fundamental challenge for ITS and is crucial for users safety. 
%Machine Learning 
\textbf{Why Radars? --}  While cameras are the by-default sensors for this task, they are generally limited under poor/degraded capturing conditions such as fog, rain, etc \cite{olimp}. Therefore, other complementary modalities such as radar technologies have been proposed in enhanced environment perception \cite{its_review,olimp}. In fact, since radars use electromagnetic waves, they are not impacted by lighting or weather conditions and challenges. %In fact, diverse ML-based methods and techniques have been developed \cite{its_review,olimp}.  %Several data modalities are exploited for this task including images, Lidar, and radar signals \cite{olimp}. 
\textbf{Why UWB Radars? --} Most radar systems are used to detect the existence, location, and trajectory of objects by analyzing the electromagnetic waves reflected by the environment. However, Ultra-Wide Band (UWB) radars have even higher utility and are particularly efficient for short range settings as they carry richer information on the environment. In fact, UWB radars deliver high-resolution signals that can be exploited to not only detect, but also recognize obstacles. This radar technology transmits very short electromagnetic pulses with low energy in the order of nanoseconds. These initial pulses are received as reflected echo signals with distortions that are directly impacted by the obstacle physical properties, and thereby represent the object signature. In fact, this signature contains information that go beyond the distance and the velocity; it is shaped by the object material, geometry and size \cite{entropy}. Due to these interesting properties, recent deep learning techniques were applied to exploiting UWB data and achieved promising environment perception results \cite{com_lett,olimp,exp}.

%, deep learning is one of the most widely used tools in this domain. Techniques such as Global Positioning System (GPS), camera based systems, infrared detectors, acoustics sensors, vibration and seismic sensors, and radar systems are used to achieve this task \cite{kocic2018sensors}. Among different radar sensors, Ultra Wide Band (UWB) radar offers high-resolution ranging in dynamic environments, low power consumption and high performance in multipath channels without requiring Line Of Sight (LOS).

In spite of Deep Neural Networks (DNNs) outstanding performance, they are shown to be vulnerable to adversarial noise \cite{pgd,fgsm,defensiveapproximation,guesmi2021sit}. Computer vision is the mainstream application that caught the attention of the community from adversarial machine learning perspective \cite{CW,fgsm,pgd}. However, the vulnerability of ML in other application domains has also been explored. For example, several papers have explored adversarial attacks on automatic speech/speaker recognition \cite{ASR,bob,asr_ufl}, Lidar \cite{lidarCVPR20} or electrocardiograms \cite{ecg_nature}. While some of the state-of-the-art work is practical in real-world conditions \cite{asr_ufl, patch,multi,face,ASR}, most of the works require very specific settings to be applicable.

Electromagnetic waves propagate in a broadcasted manner within a wireless channel that may contain a variety of radio signals. This allows a malicious actor to inject noise in the receiver side by propagating a specific adversarial noise using a rogue emitter. This makes the ML-based radio applications, a potential target for adversarial attacks that are practical under real-world conditions \cite{CCS21}. Radar systems are widely used in security-sensitive and safety critical applications, and are also vulnerable to adversarial attacks. More specifically, short range devices such as UWB radar technologies represent a highly practical attack setting because of the possibility of line-of-sight transmission conditions. Few papers in the literature target radar systems \cite{xband,fmcw}. Authors in \cite{xband} consider X-band spotlight mode radar which is utilized for hundreds of kilometers range and not practical for injecting adversarial noise due to the channel complexity. \cite{fmcw} targets short-range Frequency-modulated continuous-wave (FMCW) radars and is the closest paper to our work. However, FMCW radars give only velocity and range information, while UWB delivers a complete signature of the obstacle. An more detailed overview on the related work could be found in Section \ref{sec:related}. %, which are 

%, very few papers target these systems \cite{xband,fmcw}.

%Other applications such as in medical domain using ECG \cite{ecg_nature}have als

%%--attacks on other applications, data modalities

%%--limitations of these attacks from practicality perspective

To our knowledge, this is the first work that proposes adversarial attacks on UWB radar systems. We propose a systematic pipeline to generate robust physical adversarial examples against real-world UWB-based object detectors. Robustness is achieved in three ways:\\
\noindent \textbf{(i) Shift Robustness.} This refers to the robustness against de-synchronization. In fact, while adversarial patches location is not highly influential in some computer vision cases, we surprisingly found that even a minor shift between the initial signal and the adversarial noise crafted with state-of-the-art methods results in a low to no efficiency of the attack. Therefore, we included an aggregation of random noise locations within the noise generation to craft shift-resistant adversarial patches. \\
\noindent \textbf{(ii) Spectral Domain Robustness. } We noticed that the generated shift-resistant adversarial noise has a wide spectral signature; the generated noise contains frequency components that are beyond the expected range of an UWB radar echo. This results in a direct vulnerability against signal pre-processing defenses. To bypass these countermeasures, we clip the generated adversarial noise iteratively along with the noise magnitude budget, to keep the noise in a defined frequency range, and thereby generating shift and \textbf{fliltering-resistant} noise. \\

\noindent \textbf{(iii) Undetectability. } In addition to the magnitude budget and the frequency clipping, we also consider a limit on the noise application time, i.e., the size in time-domain of the generated noise. The motivation are behind exploring this property is that a shorter noise in time has lower risks to be observable, and hence detectable. Moreover, even with being in the same frequency domain of the victim device, adversarial noise could be detected by spectrum sensing. Shorter adversarial noise injection periods lead to lower magnitude in the spectral domain, and hence higher chances to be undetected. %\textcolor{blue}{-- to follow the last idea, we might check the magnitude of signals in the spectral domain for different noise sizes (with the same $\varepsilon$ budget of course-- } %The second is that a shorter noise in time has lower risks to be ebservable   \textcolor{red}{done, Table 9} \textcolor{blue}{Great! results follow the insight!}

%\emph{The paper will progress following these three perspectives to investigate each direction limit and gradually converge to the final approach.}  

\emph{The paper will explore each of these three perspectives to illustrate their strengths and limitations before presenting our final solution.}

\noindent \textbf{Contributions.} In summary, the contributions of this paper are as follows: 

\begin{itemize}

    \item We present a-RNA, the first attack leveraging unique characteristics of Ultra-Wide Band radar signals to inject real-world adversarial noise in a radar-based environment perception system.  
    
    \item While in simulation, the state-of-the-art attacks show high efficiency, we show that they fail under realistic conditions. First, baseline attacks are neutralized with a slight timing shift, i.e., de-synchronization between the noise and the signal. Moreover, these attacks generate noise outside the spectral range expected from the reflected echos, and can be easily defended against by pre-processing countermeasures. 
    
    \item Our approach generates adversarial noise that is: \textbf{(i) input-agnostic} to be practical in real-time, \textbf{(ii) robust to incidence delay}, i.e., to de-synchronized settings, and \textbf{(iii) filtering-robust} by tailoring adversarial noise that is in the same spectral signature of the raw signal. Therefore, a-RNA represents a practical threat to UWB-based environment perception systems. %Moreover, a-RNA generates adversarial noise that is in the same spectral signature of the raw signal, thereby evading filtering defenses by-design. 
    
    \item To further anticipate adaptive defender that uses spectrum sensing to detect a potential adversarial noise, we include a additional constraint on the noise generation mechanism on the time domain patch sizes and show that these adversarial patches are hard to detect because of their low magnitude spectral components. 
    
    \item We also analyze the robustness of our attack in the presence of adversarially trained networks. While adversarial training reduces the attack efficiency, we show that further measures need to be taken to preserve UWB radar systems integrity. % Surprisingly, we found that even adversarial training is not enough to defend against RNA. We think that this is due to the random location of the adversarial noise within the signal.  %\textcolor{red}{I need to double check this claim-- also how AT does under low size patches? I see that AT is more efficient against lower size-- is this because the adversarial samples are from this part? maybe we can discuss that the AT is harder since we have $\varepsilon~ \sim L_p $-norm space and $\varepsilon$-shifted $\sim L_p $-norm space -- }
    
    \item We open-source our codes and collected data for the community to encourage further investigations of this direction \footnote{omitted for blind review}. 
\end{itemize}

\section{Background}\label{sec:bckgnd}
\begin{figure*}[htp]
\centering
\includegraphics[width=2\columnwidth]{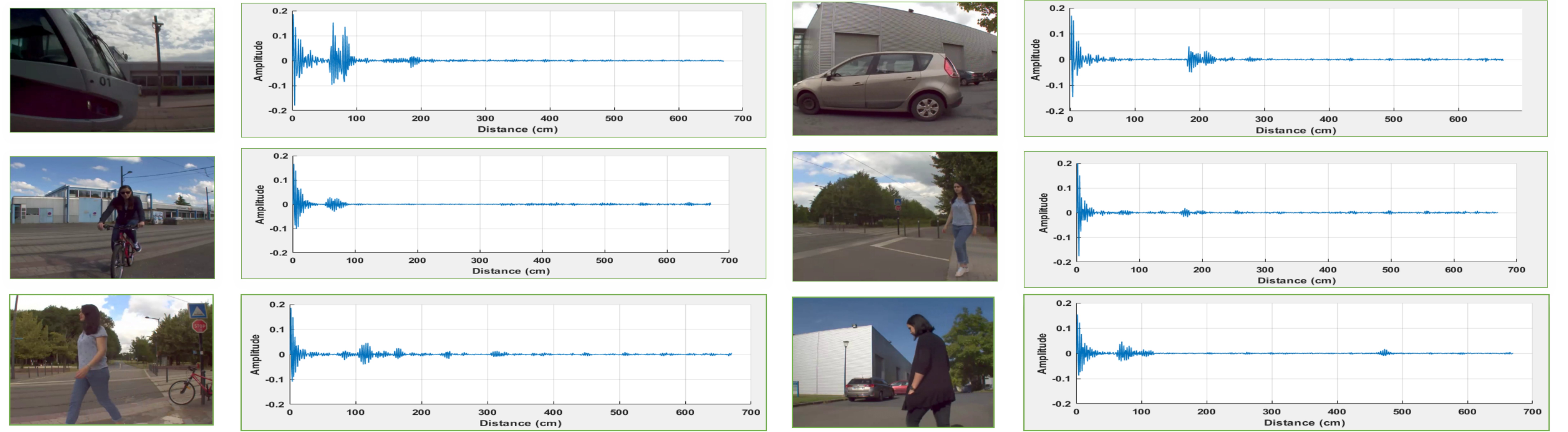} %Illustration_OLIMP.png}
\caption{Illustration of different radar signatures of different classes (with the corresponding scenes).}
\label{figure:illustration}
\end{figure*}

Ultra-wideband (UWB) is a short-range radio communication technology that allows fast and stable data transmission. UWB is generally the technology of choice for localization of moving assets in complex and space-sensitive locations due to its precision, reliability and rich information. %Due to its numerous advantages over related technologies such as RFID, BLE, or WiFi, UWB is regarded the gold standard of indoor localization technology. %It's an excellent choice for location-based automation.

The UWB transmitter emits a narrow pulse at a target's direction, and the reflected signal is detected by the UWB receiver. When a UWB pulse encounters a boundary between two types of medium with different dielectric properties during propagation, a portion of the incident electromagnetic energy is reflected back to the original medium with a reflection angle $\theta_r$ (zero reflection angle if the incident wave path is parallel to the normal line), while the other portion propagates through the next medium.
The extremely short pulses (usually in order of few nanoseconds) provide a very wide bandwidth, which has numerous advantages, including high throughput, covertness, jamming resistance, lower power, and coexistence with existing radio services \cite{Saito2003,Zetik}. UWB not only has the potential to transmit a rich data over a short distance while using very low power, but it can also pass through physical objects that tend to reflect signals with narrow bandwidth.

UWB radar technology is also employed in the automotive field \cite{wang2022, olimp} owing to the richness of information it provides, and its robustness to degraded capturing conditions. The preeminent characteristic of such technology consists in the deformation of the emitted pulse. This distortion depends on the obstacles characteristics, thereby it is labeled as the object signature. This signature is affected by the shape, material and size of the object. For example, the signature of a metallic object has a higher amplitude than that of a pedestrian. Consequently, the use of such technology remains promising for detecting objects at short range. The data acquired from the UWB radar can be represented in two forms: a one-dimensional (1D) signal, which is the reflected echo, and a two-dimensional (2D) data that can be a 2D feature map or a converted image. 

Figure \ref{figure:illustration} shows samples of UWB radar signals with corresponding images for illustration purposes. 
%\textcolor{blue}{I'll provide some examples to illustrate--}

%\textcolor{blue}{-- we might need a background section-- not everyone knows what UWB is.}

\section{Threat Model}\label{sec:threat}
%\textcolor{blue}{-- We need to present something that looks more like a threat scenario/Attack scenario rather than classical AML threat model.}

\begin{figure*}[htp]
\centering
\includegraphics[width=2\columnwidth]{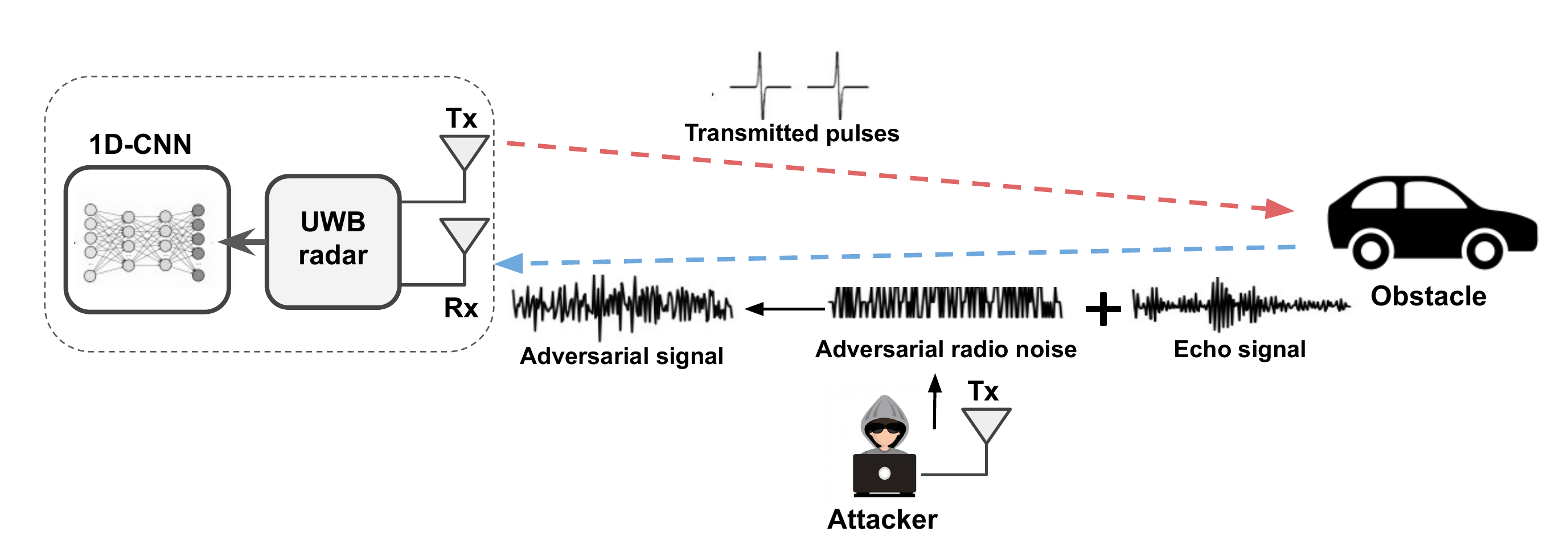}
\caption{Illustration of the threat model.}
\label{threat_model}
\end{figure*}

\subsection{Attack scenario}

An adversary wants to remotely compromise an UWB-based environment perception system such as an autonomous vehicle by causing an obstacle detection failure. 
To do so, the adversary corrupts the reflected radar signal by injecting carefully crafted adversarial radio noise in the channel.\\ 

\textbf{Physical setting. } We assume the adversary can be in the surrounding environment of the victim device. We assume that an adversary cannot physically touch the victim’s devices, alter the device settings, or install malware apps.

\textbf{Attacker knowledge. } 
We assume the adversary has access to the model, i.e., a white-box
setting. Therefore, the attacker is aware of the victim classifier's parameters and architecture. This information is used by the attacker to construct adversarial examples. 
However, at inference time, the adversaries have no access to the victim device functional parameters. They have no prior knowledge on when the system starts to send/receive the UWB signals.  

%\textbf{No third party interference. } We assume that the device is operating in a space where there is no other UWB communication is interfering within the channel.\\

\textbf{Attack equipment. } We assume that the adversaries possess wireless equipment such as a USRP and a directinal antenna that allows them to broadcast random signal in the channel and specifically in the direction of the victim device.

%We propose a new real-world adversarial radio noise attack (a-RNA) scenario. We aim to compromise the reflected signal by injecting carefully crafted adversarial radio noise. This adversarial noise is trained in the same frequency range of the raw signals making it resistant to any filtering-based defense.
%We propose white-box attacks for Radar-based Environment Perception Systems in practical settings. 

% We focus and overcome the following two important challenges that are crucial to realize the a-RNA attack on UWB radar.
% \begin{itemize}
%     \item \textbf{Input independence}: Generating adversarial perturbation based on a specific input is not possible in a practical attack scenario. For effective attacks the attacker should generate adversarial perturbation that is applicable to any possible inputs.
%     \item \textbf{Time independence}: An attacker has no prior knowledge about when the system starts to receive the echo signal. Synchronization between the adversarial perturbation and the input signal directly affects the success rate of the attack. Since most existing work generates adversarial examples for a specific input, they assume that the adversarial perturbation is perfectly synchronized with the particular input. For a reliable a-RNA, adversarial perturbations should work without needing to be perfectly synchronized with the input.
% \end{itemize}

Figure \ref{threat_model} gives an overview on our threat model in the context of environment perception setting for Intelligent Transportation Systems. An embedded UWB radar is used for obstacle detection/recognition in an autonomous vehicle. The radar signals are fed to a CNN which classifies the obstacles based on received signatures. A malicious actor injects adversarial noise in the channel to compromise the receiver side using a rogue emitter.

\subsection{Problem definition}
An adversary, using information learnt about the structure of the classifier, tries to craft perturbations added to the input to cause incorrect classification. For illustration, given an original input $x$ and a target classification model $ C(.) $, the problem of generating an adversarial example $x^*$ can be formulated as a constrained optimization \cite{pbform}:

\begin{equation}
\label{eq:adv}
     \begin{array}{rlclcl}
        x^* =  \argmin_{x^* }  \mathcal{D}(x,x^* ),\\  
         s.t.  ~ C(x^* ) = l^* ,  ~ l \neq l^* 
\end{array}
\end{equation}

Where $\mathcal{D}$ is a distance metric used to quantify similarity between two inputs (images/signals) and the goal of the optimization is to minimize the added noise, typically to avoid detection of the adversarial perturbations. $l$ and $l^*$ are the two labels of $x$ and $x^*$, respectively:  $x^*$ is considered as an adversarial example if and only if the label of the two inputs are different ($ C(x) \neq  C(x^*) $) and the added noise is bounded ($\mathcal{D}(x,x^*) < \epsilon $ where $\epsilon \geqslant 0 $).

\section{Experimental Setup}\label{sec:setup}
%---------------------
\subsection{Data Collection}
\begin{figure}[t!]
\centering
\includegraphics[width=\columnwidth]{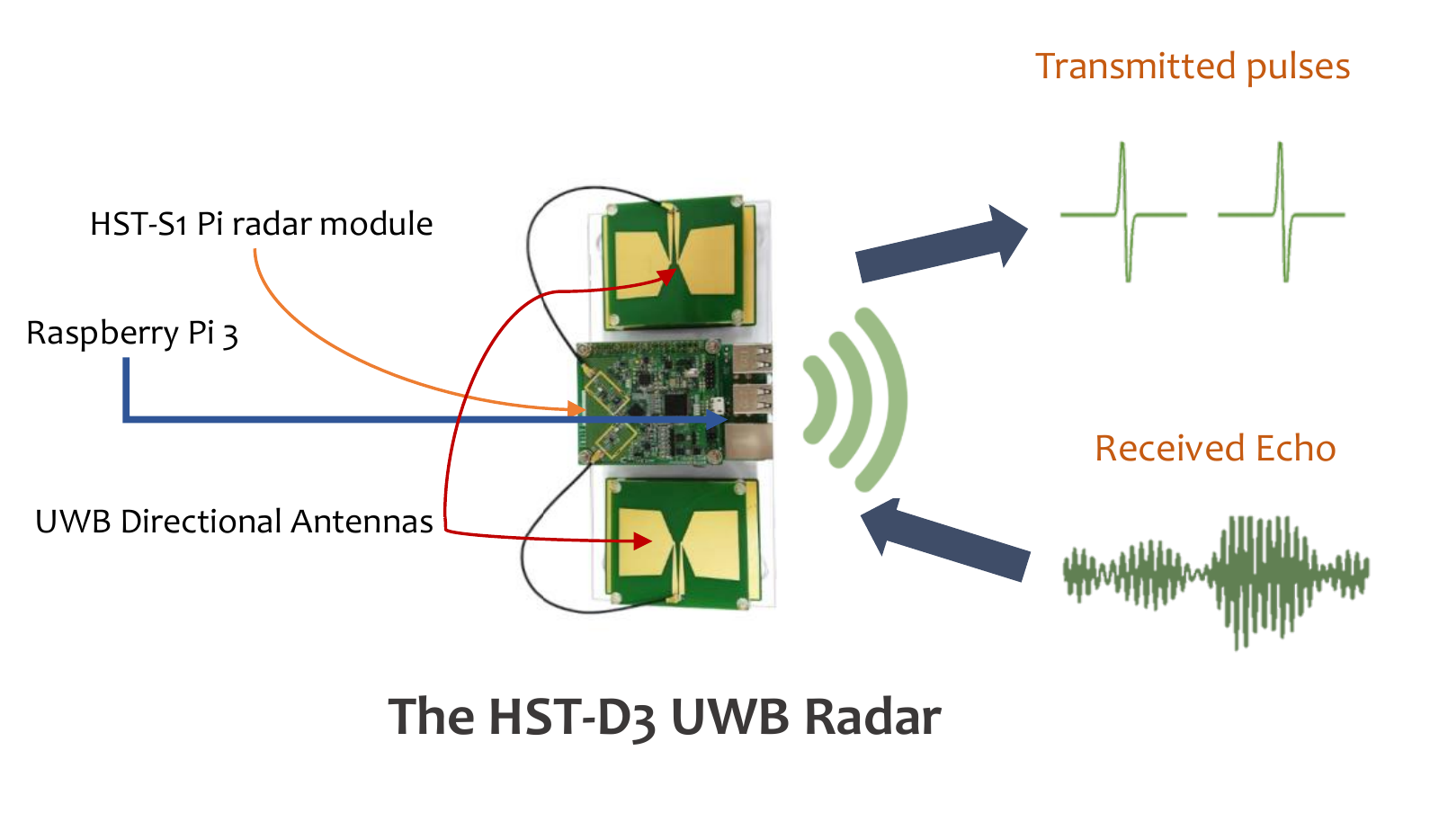}
\caption{An overview on the UWB radar system.}
\label{Fig:uwb}
\end{figure}

%---------------------
We collected UWB radar signals in real-world conditions, i.e., an outdoor environment within a University campus in which we have a variety of classes and scenarios.
The radar considered in the dataset is an UWB radar developed by the UMAIN Inc company \cite{umain} named HST-D3 with an efficient range of $6-10$ meters and a frequency range in $3 - 4$ GHz. The waveform of the generated pulse by the UWB radar is the $1\textsuperscript{st}$ order differential Gaussian pulse. We exploit the available directional antenna since it guarantees better target echo-to-clutter and noise ratio. The HST-D3 radar is a combination of HST-S1 Pi  module radar and a Raspberry Pi 3  for the acquisition. The hardware connection is presented in Figure \ref{Fig:uwb} and Table \ref{tab2} shows the radar specifications.

The dataset was captured during 3 months under different weather conditions and corresponds to 4 different classes chosen for their representativity of an urban transportation environment, namely: Car, Pedestrian, Cyclist and Tramway.

%In fact, UWB radar is robust to interference acquired from other wireless sources and its signals have high multipath immunity \cite{wang2018rf}.

\begin{table}[h]
\caption{\label{tab2} {Umain radar specifications}}
\begin{tabular}{|c|l|}
\hline
Parameter             & Value and comments   \\ \hline
Frequency range       & $3\sim4$ Ghz  \\ \hline
%Bandwidth             & 0.45$\sim$1Ghz  \\ \hline
Output Power          & Typ. -25dBm  \\ \hline
Antenna Specification & UWB Directional Antenna : \\
                      & Gain = Avg.7 dBi\\  
                      & Antenna angle (@-3dB) = \\
                      & 56°(X-Z plane) 77.5°(Y-Z plane) \\
                      & Size = 76mm x 58.5mm x 17mm \\ \hline
Sampling     & 660 samples per frame  \\ \hline
Sampling frequency    & 7,69Ghz     \\ \hline
\end{tabular}
\end{table}

%---------------------
\subsection{UWB-based System for Obstacle Detection}
%---------------------
Deep learning methods have shown great potential in solving complex problems due to their ability to automatically learn features through multiple levels of abstraction, which frees the designer from the dependence on hand-engineered features. Particularly, Convolution Neural Networks (CNNs) are very efficient in projecting raw information to learnt features' space, which represents the input space of a classifier.
More specifically, 1D-CNNs can be used to extract local 1D patches (subsequences) from uni-dimensional data samples. For obstacle recognition based on UWB signals, we consider a 1D-CNN based on 3 convolutional layers with filters of size $3x1$ and 
$2$ fully connected layers. LeakyReLU, a variant of the linear rectifier function (ReLU) is used as activation function.
 The detailed architecture hyper-parameters are presented in Table \ref{archi_1dcnn}.

\begin{table}[h]
\small
  \caption{Architecture of the 1D-CNN.}
  \label{archi_1dcnn}
  \begin{tabular}{cccc}
    \toprule
    \textbf{Layer type} & \textbf{Unit} & \textbf{Output shape}  & \textbf{\# of Parameters} \\
    \midrule
      Conv1D (LReLU) & (16,3) &  (16, 658)  &   64   \\
      Maxpool        &  (2)   &  (16, 329)  &   0      \\
      Conv1D (LReLU) & (32,3) &  (32, 327)  &   1,568  \\
      Maxpool        &  (2)   &  (32, 163)  &  0       \\
      Conv1D (LReLU) & (64,3) &  (64, 161)  &  6,208   \\
      Maxpool        &  (2)   &  (64, 80)   &  0       \\
      Flatten        &   -    &     -       & 5,120    \\     
      Linear (LReLU) &  16    &     -       &  81,936   \\
      Linear         &  5     &     -       &  85       \\ \hline
      Total parameters   &   -     &  -     &  89,861   \\
  \bottomrule
\end{tabular}
\end{table}

Training and validation samples are split in 7:3 way which gives $800$ training examples and $294$ test examples. We use a learning rate of $1e^{-4}$ and a batch size of $4$ for the training. The Adam optimizer ($\beta1=0.9$ and $\beta2=0.999$) was used with the negative log likelihood loss function. In Table \ref{acc}, we present the results of classification accuracy of the model after training for 100 epochs. %Hardware specifications: Intel(R) Xeon(R) CPU 2.20GHz.

\begin{table}[h]
\small \centering
  \caption{1D-CNN classification accuracy.}
  \label{acc}
  \begin{tabular}{cc}
    \toprule
         Classification Accuracy  & 99.31\%   \\
        % Execution time  & 2 ms   \\
  \bottomrule
\end{tabular}
\end{table}

Once we have a system with high utility, in the next sections we evaluate the vulnerability of such system to adversarial noise under realistic conditions.
%\subsection{Proposed Approach: Adversarial Wireless Noise}

%After all authors, start the affiliations section by using the {\tt \textbackslash{}affiliations} command.
%Each affiliation must be terminated by a newline {\tt \textbackslash{}\textbackslash{}} command. Make sure that you include the newline on the last affiliation too.

%\section{Adversarial Radio Noise}
%\input{sections/Proposed}

\section{Baseline Adversarial Attack on UWB systems}\label{sec:baseline}

%\begin{figure}[t!]
%\centering
%\includegraphics[width=0.9\columnwidth]{figures/accuracy_baselineA.png}
%\caption{Classification accuracy under baseline FGSM and PGD attacks.}
%\label{baseline_attacks}
%\end{figure}

%\begin{table}[h]
%\small
%  \caption{Classification Accuracy under baseline FGSM and PGD attacks for non- and adversarially trained models. \textcolor{red}{-- maybe it is better to represent in figures instead-- not sure, but we might move the AT results to a separate. section, we'll see later}} 
%  \label{baseline_attacks}
%  \begin{tabular}{ccccccc}
%    \toprule
%    \textbf{Epsilon} & \textbf{0} & \textbf{0.0005}& \textbf{0.001} &\textbf{0.0015} & \textbf{0.002}& \textbf{0.003}\\
%    \midrule
%        FGSM & 99\% & 84\% &  41\% &  16\% &  8\% &  5\% \\
%        PGD &  99\% &  77\% &  20\% &  8\% &  4\%  &  2\%  \\
        %FGSM (AT) & 99\% & 98\% &  98\% &  96\% &  96\% &  95\% \\
        %PGD (AT) &  99\% &  92\% &  92\% &  92\% &  92\%  &  92\%  \\
%  \bottomrule
%\end{tabular}
%\end{table}

%\begin{figure*}[t!]
%\centering
%\includegraphics[width=2\columnwidth]{figures/clean_adv.png}
%\caption{Clean and adversarial signals and the adversarial noise.}
%\label{Fig:time}
%\end{figure*}

In this section we proceed to a preliminary analysis of UWB-based environment perception systems vulnerability to state-of-the-art adversarial attacks. 
For this reason, we consider two situations:

\textbf{(i)} Input-specific adversarial attacks, where the adversarial noise is generated to target a specific sample. %For this case, we generate noise using two widely used attack methods, i.e., FGSM and PGD \cite{fgsm,pgd}. 

\textbf{(ii)} Universal noise that tries to alter the model output regardless of the specific sample. %, and we use the universal adversarial perturbation \cite{Universal}. 

We first present the attack generation methods, then show their corresponding results, and discuss their limits.

\subsection{Input-Specific attacks} %Attack setting. 
In this section, we build a baseline input-specific attack, where we use two state-of-the-art white-box methods, i.e., Fast Gradient Sign Method and Projected gradient descent attacks to generate adversarial examples against the UWB-based system.

\noindent Fast Gradient Sign Method (FGSM):
FGSM \cite{fgsm} is a single-step, gradient-based, attack. An adversarial example is generated by performing a one step gradient update along the direction of the sign of gradient at each element of the signal as follows:

 \begin{equation}
     x^* = x + \epsilon sign (\nabla_{x}J_{\theta}(x,y))
 \end{equation}
Where $\nabla J()$ computes the gradient of the loss function $J$ and $\theta$ is the set of model parameters. The $sign()$ denotes the sign function and $\epsilon$ is the perturbation magnitude. \\

\noindent Projected gradient descent (PGD):
PGD \cite{pgd} is a stronger iterative method where the adversarial example is generated as follows:
 \begin{equation}
x^{t+1} = \mathcal{P}_{\mathcal{S}_x}(x^t + \alpha \cdot sign (\nabla_{x}\mathcal{L}_{\theta}(x^t,y)) )
 \end{equation}
Where $\mathcal{P}_{\mathcal{S}_x}()$ is a projection operator projecting the input into the feasible region $\mathcal{S}_x$ and $\alpha$ is the added noise at each iteration.
PGD tries to find the perturbation that maximizes the loss of a model on a particular input while keeping the size of the perturbation smaller than a specified amount.\\ 

\subsection{Input-Agnostic attacks}
We also consider the universal adversarial perturbations (UAP) \cite{Universal}. UAP has been initially proposed against computer vision systems; it generates an input-agnostic adversarial patch after optimizing over a given dataset. %which are  perturbations, when added to an input signal, causes misclassification of the perturbed input.
Let $x \in \mathbb{R}^d$ be an input of dimension $d$ that follows a distribution $\mu$ ($x \sim \mu$). The main objective of a UAP is to fool a target model $C(.)$ on almost all inputs sampled from $\mu$. This problem can be formulated as finding a vector $\delta$ such that:

\begin{equation}
    C(x + \delta) \neq C(x), ~ for ~ "most"  ~ x \sim \mu
\end{equation}

Where $\delta$ represents the adversarial patch and must satisfy the following two constrains:
\begin{itemize}
    \item $\left\Vert \delta \right\Vert_p \leq \xi $
    \item $\mathbb{P}_{\substack{x \sim \mu}} \left( C(x + \delta) \neq C(x)\right) \geq 1 - \rho $
\end{itemize}

The parameter $\xi$ controls the magnitude of the perturbation vector $\delta$, and $\rho$ quantifies the desired fooling rate for all images sampled from the distribution $\mu$.

%---------------------
\subsection{Results}
%---------------------
%\textcolor{blue}{-- here come the results of the baseline-only attacks.}
We use $l_\infty$-norm as a distance metric of the noise generation. The attack success rate (defined as  1 - Classification Accuracy) represents the proportion of total perturbed signals for which the adversarial noise forces the model to output a wrong label.
Table \ref{baseline_attacks} and Table \ref{UAP} show the success rate of the input-specific and the UAP, respectively. 
As expected, these methods adapted to the UWB signals are able to generate effective adversarial examples. In the next section we discuss to which extent these results hold under realistic conditions. 

%We were also able to generate effective universal perturbations (see Table \ref{UAP}).

\begin{table}[!htp]
\small \centering
  \caption{Attack success rate of baseline attacks. } 
  \label{baseline_attacks}
  \begin{tabular}{cccccc}
    \toprule
                     &      \multicolumn{5}{c}{\textbf{Epsilon}}   \\
    \midrule
    \textbf{Attack}  & \textbf{0.001}& \textbf{0.002} & \textbf{0.005}& \textbf{0.007} &\textbf{0.01} \\
    \midrule
          PGD    &  50\% &  87\% &  96\% & 98\% &  98\%    \\
          FGSM   &  64\% &  93\% &  98\% &  98\% & 98\%   \\
  \bottomrule
\end{tabular}
\end{table}
%& \textbf{0.007} &  98\%&27\% &98\% & 22\% 
\begin{table}[!htp]
\small \centering
  \caption{Attack success rate of UAP. } 
  \label{UAP}
  \begin{tabular}{cccccc}
    \toprule
                   &           \multicolumn{5}{c}{\textbf{Epsilon}}   \\
    \midrule
    \textbf{Attack}  & \textbf{0.01} & \textbf{0.02} &\textbf{0.03} & \textbf{0.04} & \textbf{0.05}\\
    \midrule
         UAP PGD-based   &  57\% &  60\% &  70\% &  82\%  &  92\%  \\
         UAP FGSM-based &   69\% &  89\% &  90\% &  90\%  &  90\%  \\
  \bottomrule
\end{tabular}
\end{table}

%--------------
\subsection{Limits} 
%-------------
While the previous results show the vulnerability of UWB-based ML systems to adversarial attacks, the attack assumptions do not take into account the specificity of the application, nor the real-life conditions.
First, the input-specific attacks are not practical unless they are generated on-the-fly. Moreover, the adversarial noise has been applied under a perfect synchronization with the input samples, which is not practical in real-world scenarios.

\begin{figure}[tp]
\centering
\includegraphics[width=\columnwidth]{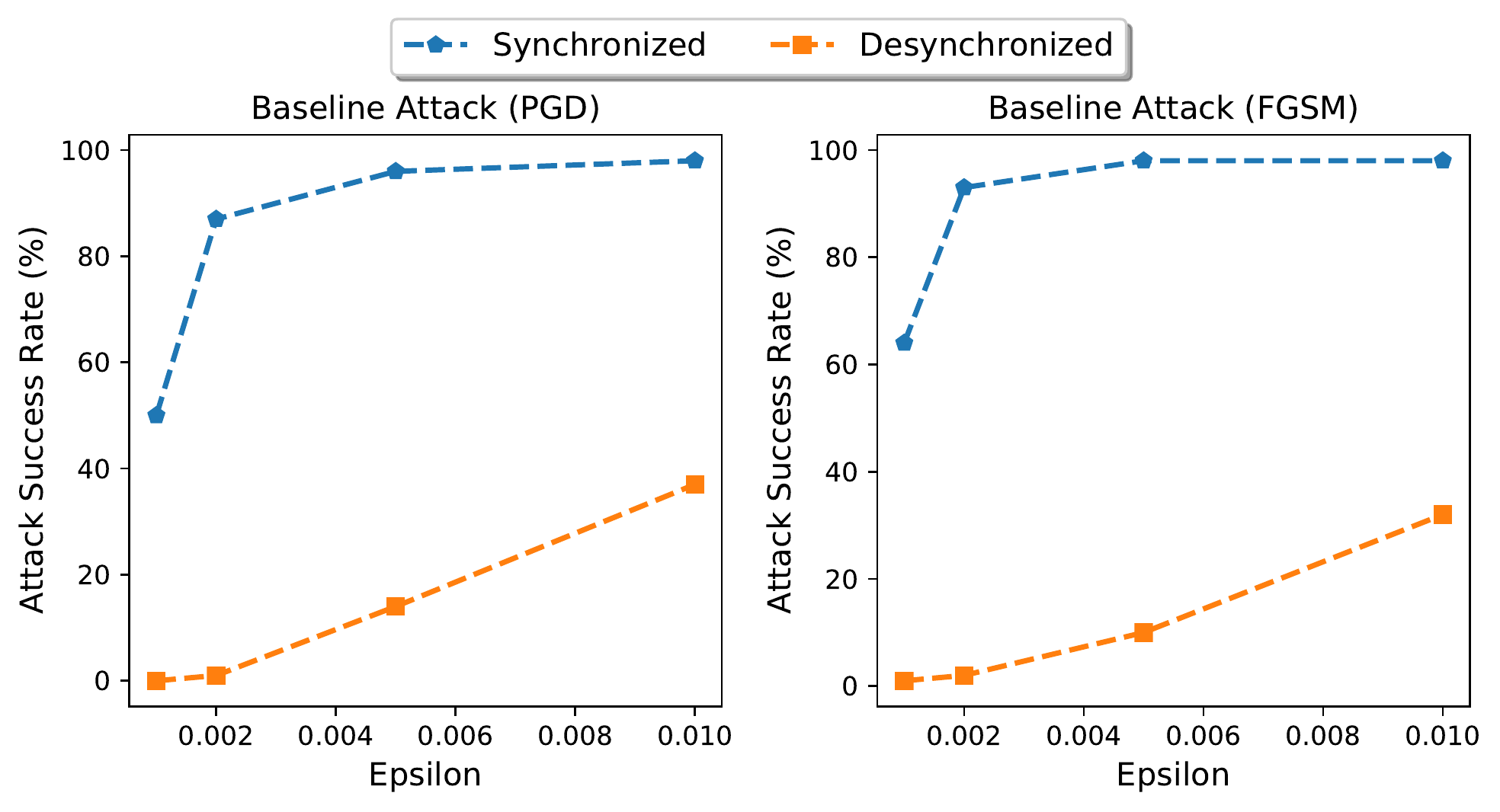}
\caption{Attack success rate of baseline attacks under synchronized and desynchronized setting.}
\label{baseline_attacks_des}
\end{figure}

In this section, we investigate the impact of de-synchronization on adversarial attacks success rate.
We apply random \emph{time shifts} to the noise incidence, which correspond to the delays in the adversarial noise incidence at receiver side. The results are depicted in Figure \ref{baseline_attacks_des}; we notice a huge drop in attacks effectiveness. In fact, the baseline attacks with small noise magnitudes are totally neutralized if not synchronized with the input. For instance, we noticed a drop from $87\%$ to $1\%$ for a noise magnitude $epsilon = 0.002$ and for higher noise budget constraints, we notice a drop of at least $61\%$ in success rate. 

\begin{figure}[tp]
\centering
\includegraphics[width=\columnwidth]{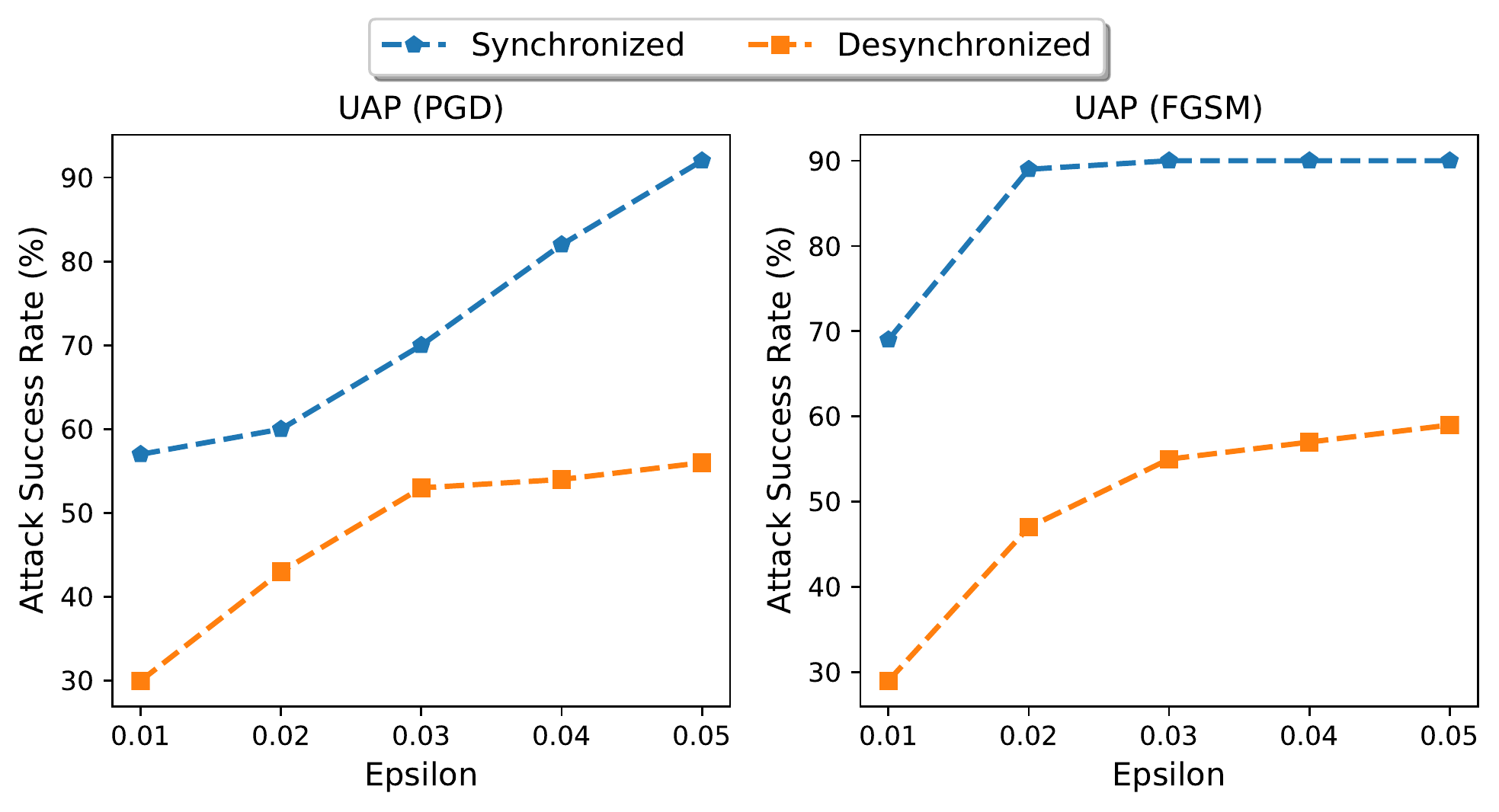}
\caption{Attack success rate of UAP under synchronized and de-synchronized adversarial settings.}
\label{UAP_des}
\end{figure}

The same trend has been shown by UAP as shown in Figure \ref{UAP_des}; we notice considerably lower attack success rates. With a noise budget equal to $0.05$ the attack is $36\%$ less efficient.
These attacks require perfect synchronization in order to be effective which is not practical since usually it's extremely hard for the attacker to predict the exact time the system starts to receive the signal. 

%Input specific attacks require real-time noise generation to be practical since the target samples are discovered on the fly. Therefore, in the remainder of the paper we consider exclusively adversarial patches that are universal, i.e., input-agnostic. 

%are performed under real-time constraints. Rapid (real-time) adversarial examples generation is crucial \cite{guesmi2022room}, which is impractical and unrealizable since providing a desync-resistant attack requires offline pre-processing for each input sample.

\begin{tcolorbox}

\noindent
\textbf{\textit{Terminology.}} Henceforth, we consider universal adversarial noise only, and we use adversarial \emph{patch} and adversarial \emph{noise} interchangeably.

\end{tcolorbox}

\section{Shift Resistant Patch (SRP)}\label{sec:srp}
In this section, we propose Shift Resistant Patch (SRP), an adversarial adversarial patch generation methods that overcomes desynchronization with the target sample.
%works for most inputs and possible shifts. %SRP overcomes the input- and time-independence challenges. 

%---------------------
\subsection{Approach}
%---------------------
As previously shown, a slight timing shift that de-synchronizes the adversarial noise with the input neutralizes the attack.
To overcome this limitation, we propose to generate adversarial noise that can efficiently fool the classifier regardless of the noise incidence delay. For this reason, we include the noise incidence time within the optimization problem. Specifically, we iteratively update the adversarial noise by aggregating its impact over different incidence times. % using \nabla_{x+k} instead of \nabla_{x}, where k models the . 

Our main objective is to fool a victim model $C(.)$ on almost all the input signals from a distribution $\mu$ in $\mathbb{R}^d$ \emph{whenever} the adversarial noise incidence time within the victim signal. This problem can be formulated as finding a vector $\delta$ such that:

\begin{equation}\label{eq:srp}
\begin{split}
        C(x + shift(\delta, k)) \neq C(x), ~ for ~ most ~ x \sim \mu \\ s.t. \left\Vert \delta \right\Vert_p \leq \epsilon 
\end{split}
\end{equation}

Where the parameter $\epsilon$ controls the magnitude of the noise vector $\delta$ and 
$shift(.)$ is a function that quantifies the adversarial patch signal $\delta$, relatively with regard to a target signal $x$ given a time incidence $k \in [0, d] $. Given an adversarial patch  $\delta=\{\delta_j\} \forall j \in[0, d] $ that is \emph{repeatedly broadcasted in a continuous loop} within the channel, the function $shift(.)$ could be expressed as:\\

\begin{equation}\label{eq:shift}
\begin{split}
shift(\delta, k)=\left\{\begin{array}{lll}
\delta(d-k+j) & \text{if} & j \in[0, k] \\
\delta(j-k) & ~ \text{else}
\end{array}\right.\\
\end{split}
\end{equation}

%\textcolor{blue}{Detailed explanation of the algorithm needed--\\}
In Algorithm \ref{SRP}, we iterate across the data in the batch $X$ gradually updating the adversarial patch. %At each iteration, the perturbation update $\Delta \delta_i$ is computed using $attack()$ function for a randomly chosen locations $k$. 
For each input example we generate the corresponding adversarial noise using $attack()$ function for a randomly chosen locations $k$.
The function $attack()$ is detailed in Algorithm \ref{attack_alg1}; it is a gradient-based attack that performs $m$ steps along the direction of the sign of gradient at each element of the signal in a way that maximizes the loss of a model. Therefore, to update the patch every iteration, we use $\nabla_{x+ shift(\delta,k)}(.)$ instead of using $\nabla_{x+\delta}(.)$. If the generated adversarial example $x^{adv}_i$ fools the model, we undo the shifting and rearrange the adversarial noise back to the original form (before performing the $shift$). This noise is aggregated to the current instance of the patch and than projected on the $L_p$ norm ball of size $\epsilon$.

\begin{algorithm}[!t]
\caption{Shift Resistant Patch (SRP).}
\label{SRP}
\begin{algorithmic}[1]
\State \textbf{Input:} Data batch: $X$, classifier: $C$, noise magnitude: $\epsilon$, step size: $\alpha$, number of iterations: $N$, raw signal size: $d$.

\State \textbf{Output:} $\delta$ adversarial patch.

\State Initialize $\delta \leftarrow 0$
\While{$iter < N$ }
    \For{each data point $x_i \in X$}
        \If{$C(x_i + shift(\delta, k)) == C(x_i)$}
            \State Select a random location $k$
            %\State $\Delta\delta_i \leftarrow  attack(C, x_i + shift(\delta, k), y_i, \epsilon, \alpha)$ 
            \State $x^{adv}_i \leftarrow  attack(C, x_i + shift(\delta, k), y_i, \epsilon, \alpha)$ 
            \If{$C(x_i) != C(x^{adv}_i)$}
                \State $\Delta\delta_i \leftarrow concat((x^{adv}_i - x_i)[k:d], (x^{adv}_i - x_i)[0:k])$
                \State $\delta \leftarrow \mathcal{P}_{l_p, \epsilon}(\delta + \Delta \delta_i)$
\EndIf
\EndIf
\EndFor
\State $iter += 1$
\EndWhile
%\textcolor{red}{-- please double check the indexes not to be confused with vector elements-- }
\Function{$shift$}{$\delta$, $k$}
\State Initialize $shifted\_\delta \leftarrow 0$
\State $shifted\_\delta \leftarrow concat(\delta [d-k:d],\delta [0:d-k] )$

\EndFunction

\end{algorithmic}
\end{algorithm}
%--------------------------------------------------
\begin{algorithm}[!tp]
\caption{$attack$ function.}
\label{attack_alg1}
\begin{algorithmic}[1]
\State \textbf{Input:} a classifier: $C$ with loss $J$, noise budget:  $\epsilon$, step size: $\alpha$,input signal: $x$, label: $y$, number of iterations: $m$.
\State \textbf{Output:} $x^{adv}$
%\State Initialize~ $\delta \leftarrow 0$
\State Initialize~ $x^{adv} \leftarrow 0$

\For {i= 0...m-1}
    %\State prediction = $argmax(C(x_{adv})) $
    \State ${x_{i+1}}^{adv} = Clip \{x_i + \alpha sign(\nabla_{x_{i}}^{adv}J_{\theta}(C({x_{i}}^{adv}),y))\}$
\EndFor

%\State $\delta = x^{adv} - x$
\end{algorithmic}
\end{algorithm}
%----------------------------------------
%\begin{algorithm}
%\caption{$shift$ function.}
%\label{shift}
%\begin{algorithmic}[1]
%\State \textbf{Input:} $\delta$ noise, $k$ a random location, $maxlen$ input signal length.
%\State \textbf{Output:} $shifted\_\delta$ noise starting from position $k$.
%\State Initialize $shifted\_\delta \leftarrow 0$
%\State $shifted\_\delta \leftarrow concat(\delta [ 0,0,k:maxlen ],\delta [ 0,0,0:k ] )$
%\end{algorithmic}
%\end{algorithm}
%-------------------------------------------------
%----------------
\subsection{Results}
%----------------
In this section, we evaluate SRP under random noise incidence delays. We generate adversarial noise using Algorithm \ref{SRP}, and evaluate the attack success rate for different noise magnitudes. The evaluation is performed  comparatively with  to UAP under random incidence time of the adversarial noise. % and measure the success rate of each attack. After crafting the adversarial patches we test them under a desynchronized setting.

\begin{figure}[htp]
\centering
\includegraphics[width=\columnwidth]{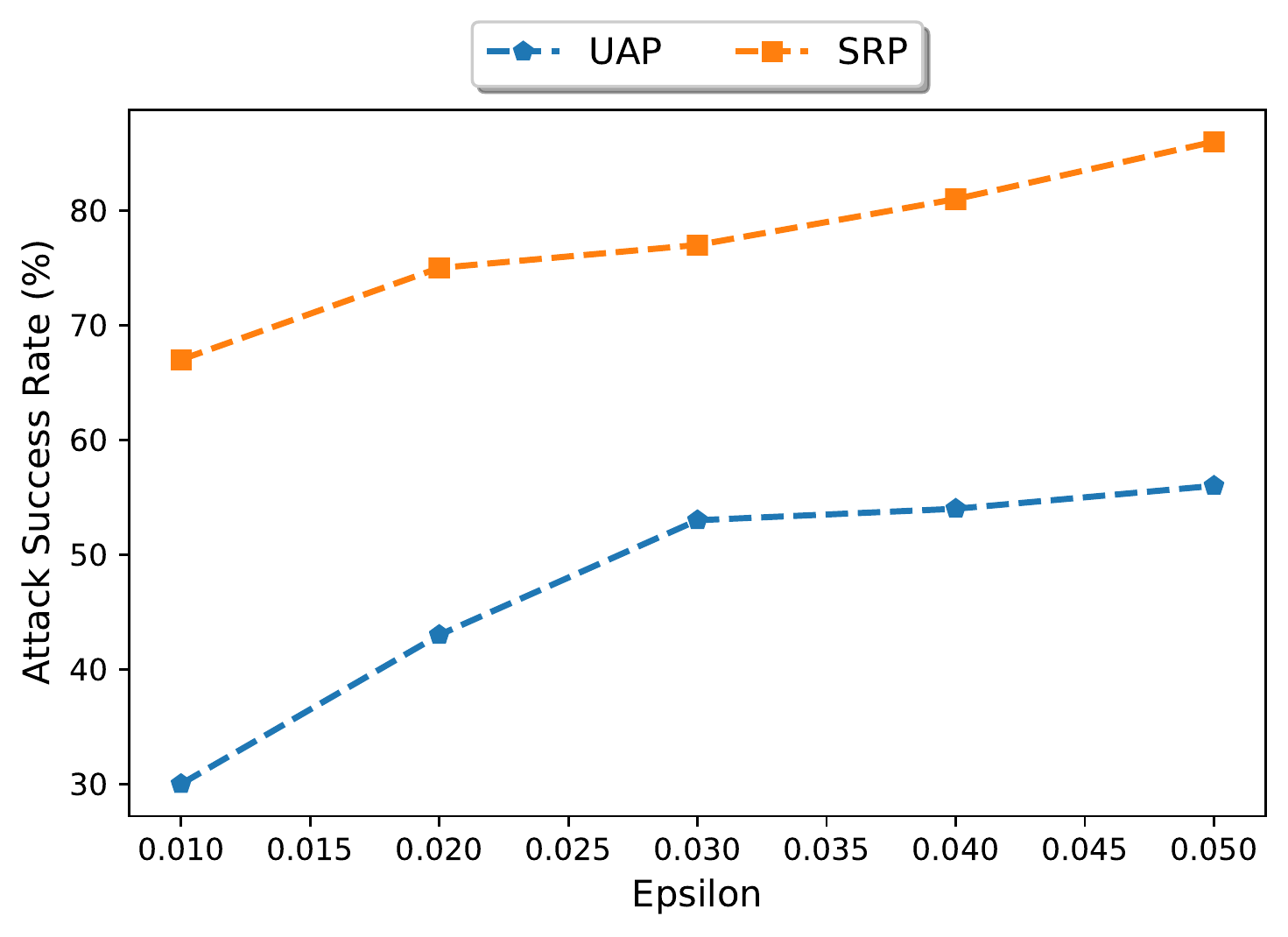}
\caption{Attack success rate of proposed SRP compared to UAP.}
\label{srp}
\end{figure}

As shown in Figure \ref{srp}, SRP is able to recover the adversarial attack effectiveness. For instance SRP has $30\%$ higher success rate than UAP for a noise budget equal to $0.05$. %SRP outperforms UAP in being robust against change in injection time.

%\begin{table}[h]
%\small \centering
  %\caption{Attack success rate of the proposed SRP vs UAP under desynchronized setting. %\textcolor{blue}{Probably a figure is better--}} 
%  \label{proposed_alg1}
%  \begin{tabular}{cccccc}
%    \toprule
%                     & \multicolumn{4}{c}{\textbf{Epsilon}}   \\
%    \midrule
%    \textbf{Attack}  & \textbf{0.01}& \textbf{0.02} &\textbf{0.03} & \textbf{0.04}& \textbf{0.05}\\
%    \midrule
%        UAP        &   30\% &  43\% &  53\% &  54\%  & 56 \%  \\
%        %UAP (FGSM-based)      &  29\% &  47\% &  55\% &  57\%  &  59\%  \\
%        SRP                 &  67\% &  75\% &  77\% &  81\%  &  86\%  \\
%        %Proposed FGSM-based &  40\% &  70\% &  81\% &  85\%  &  88\%  \\
%  \bottomrule
%\end{tabular}
%\end{table}

\subsection{Limits}

While the proposed attack addresses the synchronization limit, the threat model assumes a baseline victim device without any countermeasures. However, one of the straightforward protection of such systems is to pre-process the input to filter out undesired signals from non relevant frequency ranges. 

\subsubsection{Impact of a pre-processing defense}
%---------------------
In this section, we consider a system that is comprehensively pre-processing the input signal. The defender defines its target frequency range based on the expected echos from the environment. We evaluate the efficiency of SRP under this setting for different noise magnitudes.

The frequency spectrum is obtained using the Fast Fourier Transform (FFT), which is defined by
\begin{equation}
    F(k) = \sum^{N-1}_{n=0}f(n)e^{-i\frac{2\pi}{N}nk}
\end{equation}
Where $N$ is the length of the spectral signature and $f$ is the original time domain signal.

For each window, an FFT generates a frequency domain representation of the signal referred to as magnitude spectrum. The magnitude spectrum details each frequency and the corresponding intensity that make up a signal.

The Inverse Fast Fourier Transform (IFFT) converts frequency domain signal to time domain signal and is expressed as follows%. IFFT computes the one dimensional inverse discrete Fourier transform of input.

\begin{equation}
    f(n) = \frac{1}{N} \times \sum^{N-1}_{n=0}F(k)e^{i\frac{2\pi}{N}nk}
\end{equation}

Where $F(k)$ is the frequency domain magnitudes and $f(n)$ is the recovered time domain samples.

%\textcolor{blue}{ Figures \ref{filtered_uap} and \ref{filtered_srp}: I suggest to also add for both of them the illustration of time domain signals before and after filters-- } \textcolor{red}{Do you think I should add the raw signal to figures 8, 10 and 13?}-- No I don't think it adds something. 

\begin{figure*}[!ht]
\centering
\includegraphics[width=2\columnwidth]{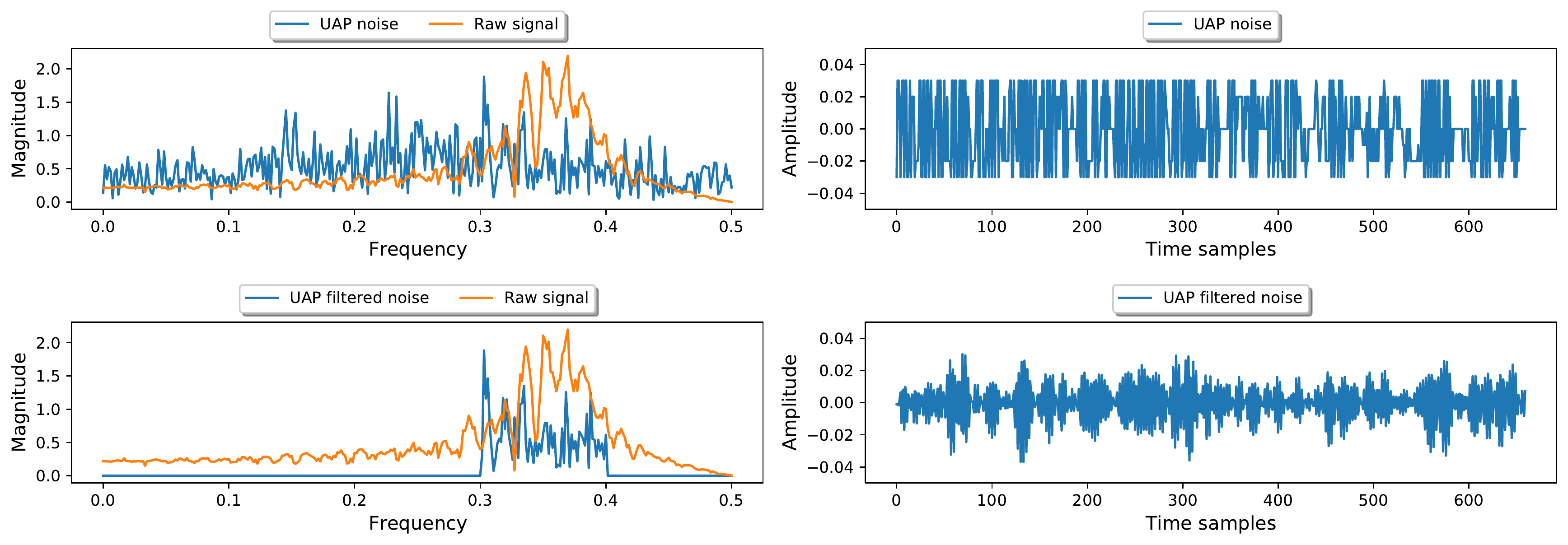}
\caption{UAP adversarial noise before and after filtering in the spectral domain (left) and the time domain (right). %\textcolor{blue}{Please notce: blue corresponds to noise on the right and to signal in the left-- }
}
\label{filtered_uap}
\end{figure*}

\begin{figure*}[!ht]
\centering
\includegraphics[width=2\columnwidth]{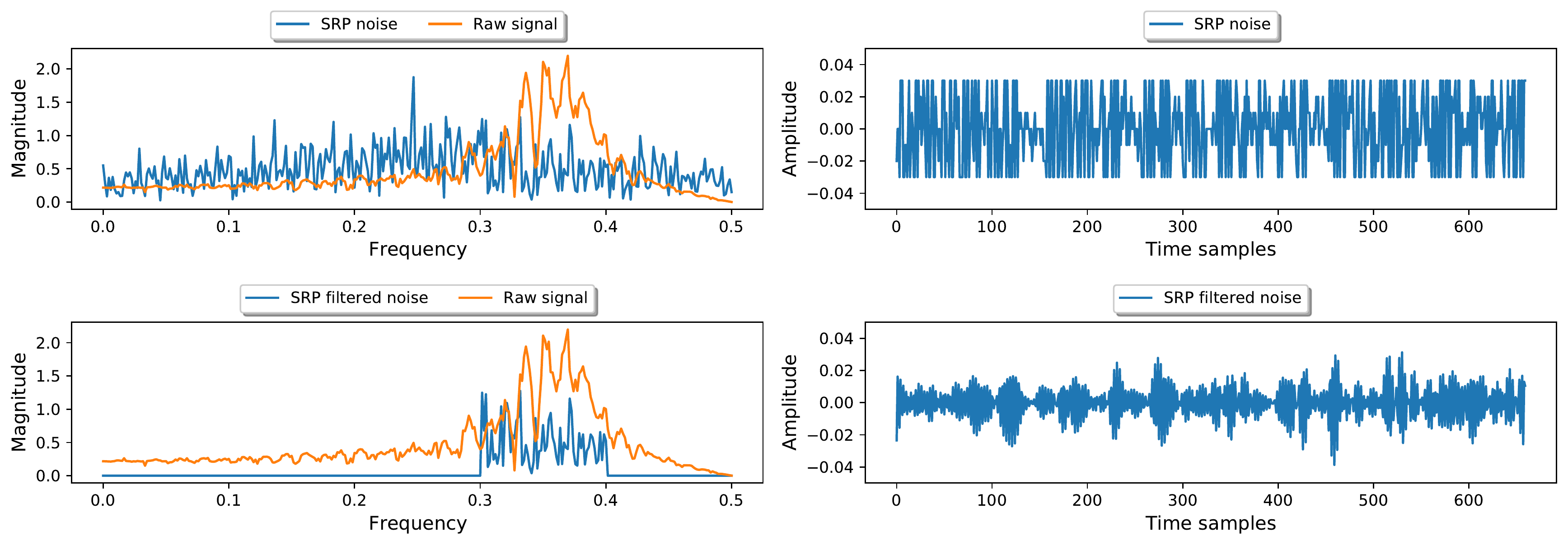}
\caption{SRP adversarial noise before and after filtering in the spectral domain (left) and the time domain (right).}
\label{filtered_srp}
\end{figure*}

%\begin{figure}[htp]
%\centering
%\includegraphics[width=\columnwidth]{figures/uap_time domain.png}
%\caption{Time domain of UAP adversarial noise before and after the frequential mask.}
%\label{time_uap}
%\end{figure}

We first explore the power spectral density of the raw data to identify the frequency range that represents the region of interest. Specifically, we identify the averege minimum and maximum frequencies ($[f_{min}, f_{max}]$ that contain $95\%$ of the total spectral power. Accordingly, we propose to use a pass-band filter corresponding to this frequency range, and make sure that there is no baseline accuracy drop is noticed under this pre-processing. 

Algorithm \ref{filter} gives an overview on the pre-processing procedure. We first transfer the signal to the spectral domain using the FFT. Then we apply a low-pass filter using $f_{min}$ as a cut-off frequency, followed by a high-pass using $f_{max}$ as cut-off. 
To finish, we recover the time domain signal using the IFFT. 
The resulting procedure represents a clipping operation in the spectral domain that we will use later in Section \ref{sec:sfr} for an adaptive attack.  %of  signal to a specific frequency range.

Figures \ref{filtered_uap} and \ref{filtered_srp} illustrate the frequency spectrum of the UAP and SRP-generated adversarial noise compared to the raw signal. Since the generation process of SRP has no restriction on the frequency, we notice a wide spectral signature, i.e., the generated noise contains frequency components that are beyond the expected range of an original raw signal which makes it vulnerable to filtering-based defenses.

Figures \ref{filtered_uap} and \ref{filtered_srp} also show the impact of the filter on the time domain signal for UAP and SRP-generated noise. While the filtered noise signals are significantly changed, we confirm this by studying their adversarial impact post-filtering.

\begin{algorithm}
\caption{$filter$: Pass-band Filter.}
\label{filter}
\begin{algorithmic}[1]
\State \textbf{Input:} time domain signal: $x$, Min Frequency: $f_{min}$ , Max Frequency: $f_{max}$ 
\State \textbf{Output:} $filtered\_x$
\State Initialize~ $filtered\_x \leftarrow 0$
\State $X \leftarrow FFT(x) $
\State $X \leftarrow LowPass(X, f_{min})$
\State $X \leftarrow HighPass(X, f_{max})$
%\State $X \leftarrow pass1 \times pass2 $
\State $filtered\_x \leftarrow IFFT(X)$
\end{algorithmic}
\end{algorithm}

%\begin{figure}[htp]
%\centering
%\includegraphics[width=\columnwidth]{figures/srp_time domain.png}
%\caption{Time domain of SRP adversarial noise before and after the frequential mask.}
%\label{time_srp}
%\end{figure}

%\subsubsection{Input-Specific attacks}
%When evaluating the attack success rate we notice a drop of $68\%$
%\begin{table}[h]
%\small \centering
%  \caption{Attack success rate of baseline PGD and FGSM attacks with and without frequency range restriction (FRR).}
%  \label{filters_IS}
%  \begin{tabular}{ccccccc}
%    \toprule
%                        &          & \multicolumn{5}{c}{\textbf{Epsilon}}   \\
%    \midrule
%    \textbf{Attack} & \textbf{FRR} & \textbf{0.0005} & \textbf{0.001} & \textbf{0.0015} & \textbf{0.002} & \textbf{0.003} \\
%    \midrule
%        PGD     & without  & 23\% &  80\%  &  92\% &   96\% & 98\% \\
%                & with     & 2\%  &  12\%  &  21\% &   24\% & 30\% \\
%        FGSM    & without  & 16\% &  59\%  &  84\% &   92\% & 95\% \\
%                & with     & 1\%  &  6\%   &  17\% &   30\% & 49\% \\
%  \bottomrule
%\end{tabular}
%\end{table}

%---------------------------------
%\subsubsection{Input-Agnostic attacks}
%---------------------------------
%Same trend was noticed with universal attacks (input-agnostic attacks).

%We use the pass-band filter to limit the frequency range of generated patches to $[0.3, 0.4]$ (the main frequency range of the raw signal) as shown in Figure \ref{filtered_uap} and \ref{filtered_srp}. 

\begin{figure}[t]
\centering
\includegraphics[width=\columnwidth]{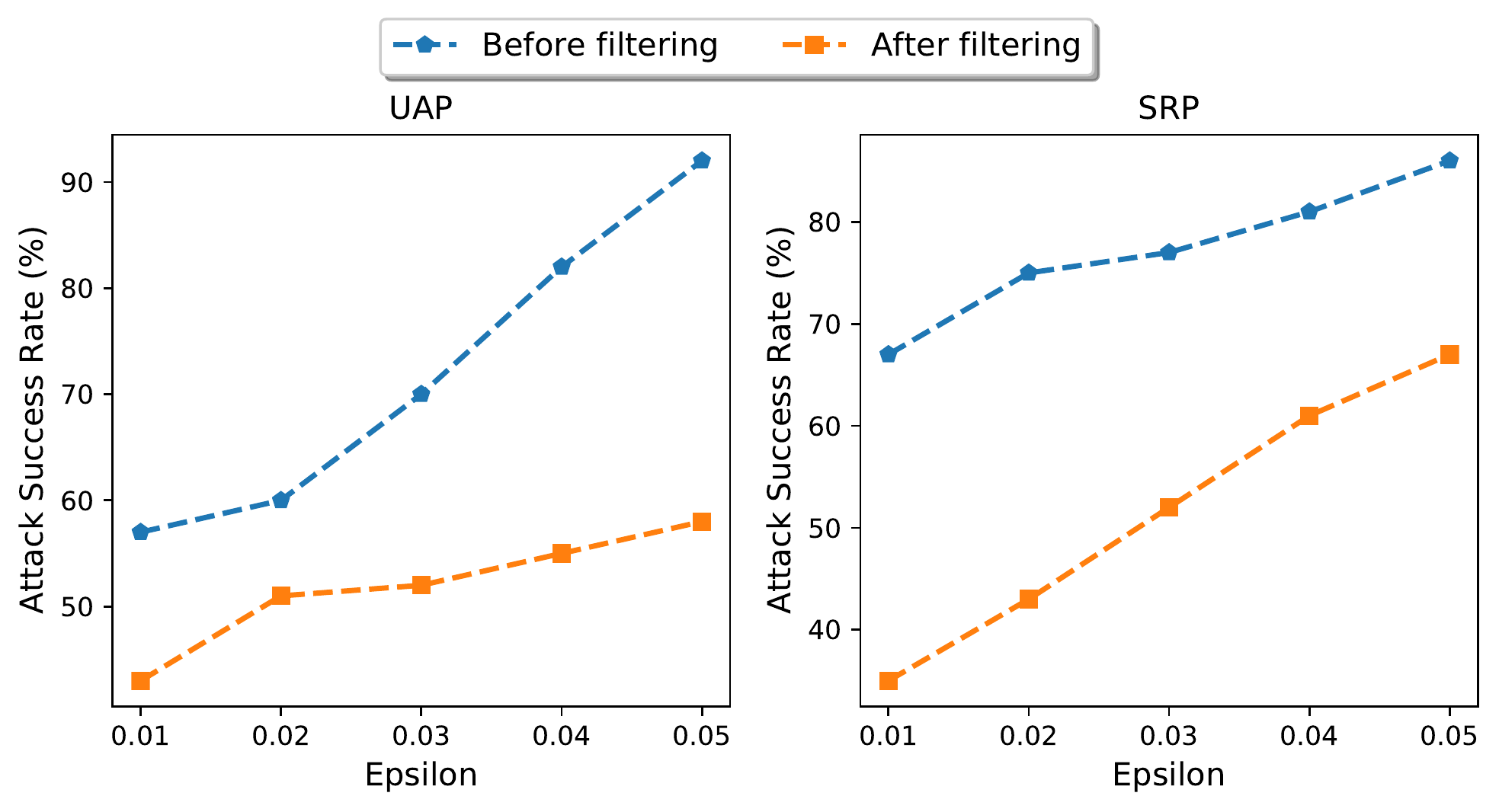}
\caption{Attack success rate of UAP and SRP under passband filter.}
\label{filters_UAP}
\end{figure}

Figure \ref{filters_UAP} presents the attack success rate of UAP and SRP. As a matter of fact, the filter significantly degrades the performance of both attack methods. For UAP, we notice a drop of $34\%$ for a noise budget of $0.05$, while SRP-generated noise (Algorithm \ref{SRP}) is $20\%$ less effective when limiting the frequency range. 

The next section proposes an adaptive attack to this defense mechanism. 

% \begin{figure*}[htp]
% \centering
% \includegraphics[width=2\columnwidth]{figures/srp_filtered.pdf}
% \caption{Frequency spectrum of SRP adversarial noise before and after the frequential mask.}
% \label{filtered_srp}
% \end{figure*}

%\begin{table}[h]
%\small \centering
%  \caption{Attack success rate of UAP (PGD-based) and SRP with and without frequency range restriction (FRR).} 
%  \label{filters_UAP}
%  \begin{tabular}{ccccccc}
%    \toprule
%                        &          & \multicolumn{5}{c}{\textbf{Epsilon}}   \\
%    \midrule
%    \textbf{Attack} & \textbf{FRR}  & \textbf{0.01} & \textbf{0.02} & \textbf{0.03} & \textbf{0.04}& \textbf{0.05}\\
%    \midrule
%        UAP        & without & 57\% &  60\%  &  70\% &   82\%  & 92\% \\
%                    & with    & 43\% &  51\%  &  52\% &   55\%  & 58\% \\
%        %UAP (FGSM-based) & without &  69\% &  89\% &  90\% &  90\%  &  90\% \\
%        %                 & with    &  28\% &  41\%  &  47\% &  66\% & 77\% \\ 
%        SRP        & without & 67\% &  75\%  &  77\% &   81\%  & 86\%  \\
%                    & with    & 35\% &  43\%  &  52\% &   61\%  & 67\% \\
%        %Proposed (alg1) (S)  & with  & 39\%  &  54\%  &  62\% &   68\%  & 78\% \\
%        %FGSM (alg1) & without & 40\% &  70\% &  81\% &  85\%  &  88\%  \\
%        %            & with    & 12\% &  48\%  &  49\% &   44\%  & 43\% \\
%  \bottomrule
%\end{tabular}
%\end{table}

\section{Shift \& Filtering Resistant Patch (SFR)}\label{sec:sfr}
In this section we propose  SFR (Shift \& Filtering Resistant Patch) to generate adversarial noise that is robust against incidence delay and additionally adaptive to filtering defense.

    \subsection{Approach}
The objective of SFR is to generate adversarial noise in a specific frequency range which corresponds to the expected range of the radar echoes. Consequently, the defender will not be able to cut-off the impact of the injected noise unless with a loss of utility. For this reason, SFR not only clips the noise magnitude in the time domain to fit a noise budget, but also adds a new constraint on the noise generation in the spectral domain. In fact, SFR also projects the noise in a subset of the frequency domain to restrain its spectral components. 
The problem can therefore be formulated as follows:

\begin{equation}\begin{split}
        C(x + shift(\delta, k))) \neq C(x), ~ for ~ most ~ x \sim \mu ~ ~ s.t.\\ 
        \left\Vert \delta \right\Vert_p \leq \epsilon \\
         ~FFT(\delta) \in [f_{min}, f_{max}]\\
\end{split}
\end{equation}

Where $\epsilon$ controls the magnitude of the noise vector $\delta$ and 
$shift(.)$ is a function expressed by Equation \ref{eq:shift} that quantifies the adversarial patch signal $\delta$, relatively with regard to a target signal $x$ given a time incidence $k \in [0, d] $.
The new constraint on $\delta$ limits the noise frequency components to an acceptable range defined by $f_{min}$ and $f_{max}$.

Algorithm \ref{patch_fft} details the noise generation mechanism. In SFR, we include the filtering step within the adversarial noise generation procedure.  For each iteration, after updating the noise using $attack()$ function, we project the noise back to the target spectral domain using the pass-band $filter$ defined previously. Therefore, we only retain noise samples with the same frequency range as the raw signal, these samples are then aggregated to the current patch instance.

\begin{algorithm}[t]
\caption{Shift \& Filtering Resistant Patch (SFR).}
\label{patch_fft}
\begin{algorithmic}[1]
\State \textbf{Input:} Data batch: $X$, classifier: $C$, noise magnitude: $\epsilon$, number of iterations:  $N$, min frequency: $f_{min}$, max frequency: $f_{max}$ .
\State \textbf{Output:} $\delta$ trained patch.

\State Initialize $\delta \leftarrow 0$
\While{$iter < N$ }
    \For{each data point $x_i \in X$}
        \If{$C(x_i + shift(\delta, k)) = C(x_i)$}
            \State Select a random location $k$
            \State $x^{adv}_i \leftarrow  attack(C, x_i + shift(\delta, k), y_i, \epsilon, \alpha)$ 
            \If{$C(x_i) != C(x^{adv}_i)$}
                \State $\Delta\delta_i \leftarrow concat((x^{adv}_i - x_i)[k:d], (x^{adv}_i - x_i)[0:k])$
                \State $\delta \leftarrow \delta + \Delta \delta_i$
                %\State $fft\_\delta \leftarrow FFT(\delta)$
                %\State $\delta \leftarrow IFFT(fft\_\delta \times kernel)$
                \State $\delta \leftarrow filter(\delta, f_{min}, f_{max} )$
                \State $\delta \leftarrow \mathcal{P}_{l_p, \epsilon}(\delta)$
\EndIf
\EndIf
\EndFor
\State $iter += 1$
\EndWhile
\end{algorithmic}
\end{algorithm}
%---------------------

\subsection{Results}

\begin{figure*}[htp]
\centering
\includegraphics[width= 2\columnwidth]{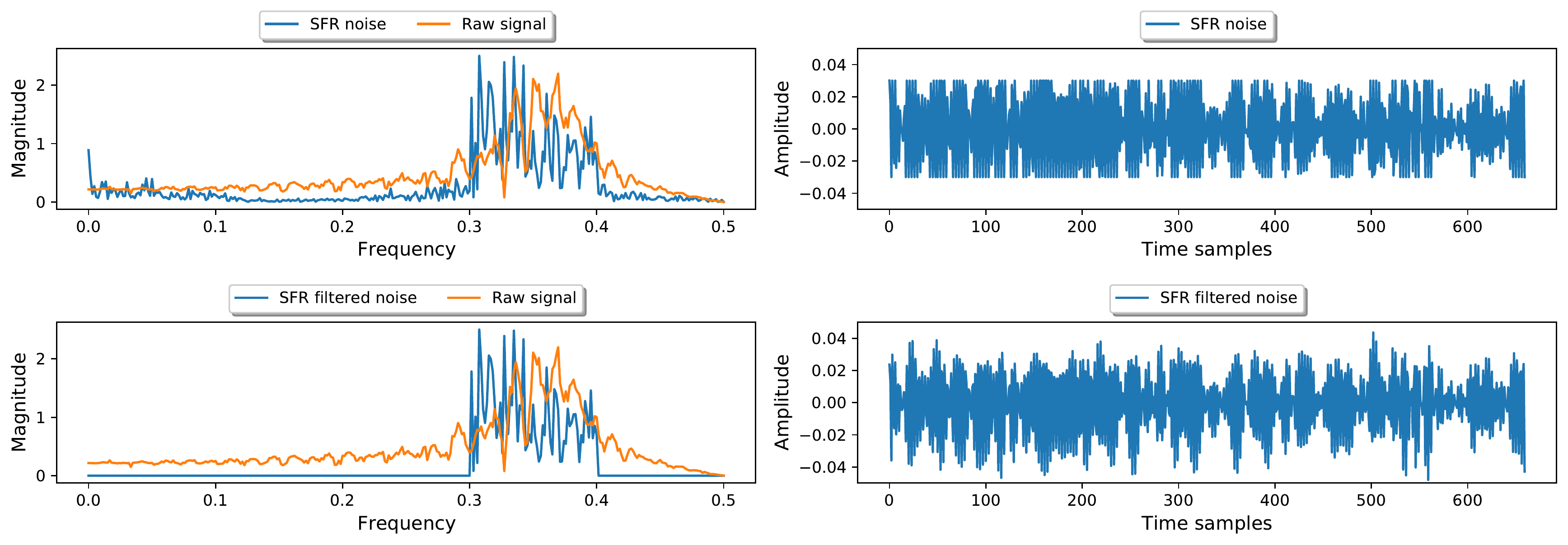}
\caption{Frequency spectrum of SFR adversarial noise.}
\label{filtered_sfr}
\end{figure*}

%\begin{figure}[htp]
%\centering
%\includegraphics[width=\columnwidth]{figures/sfr_time_domain.png}
%\caption{Time domain of SFR adversarial noise.}
%\label{time_sfr}
%\end{figure}

Figure \ref{filtered_sfr} gives an illustration of the filtering impact on a SFR-generated noise both in the frequency and time domains. In contrast with the previous results, the illustration shows that SFR-generated noise is not substantially impacted by the filter. %Using SFR, we were able to generate patches that satify both noise magnitude and frequency range constraints.

\begin{figure}[htp]
\centering
\includegraphics[width= \columnwidth]{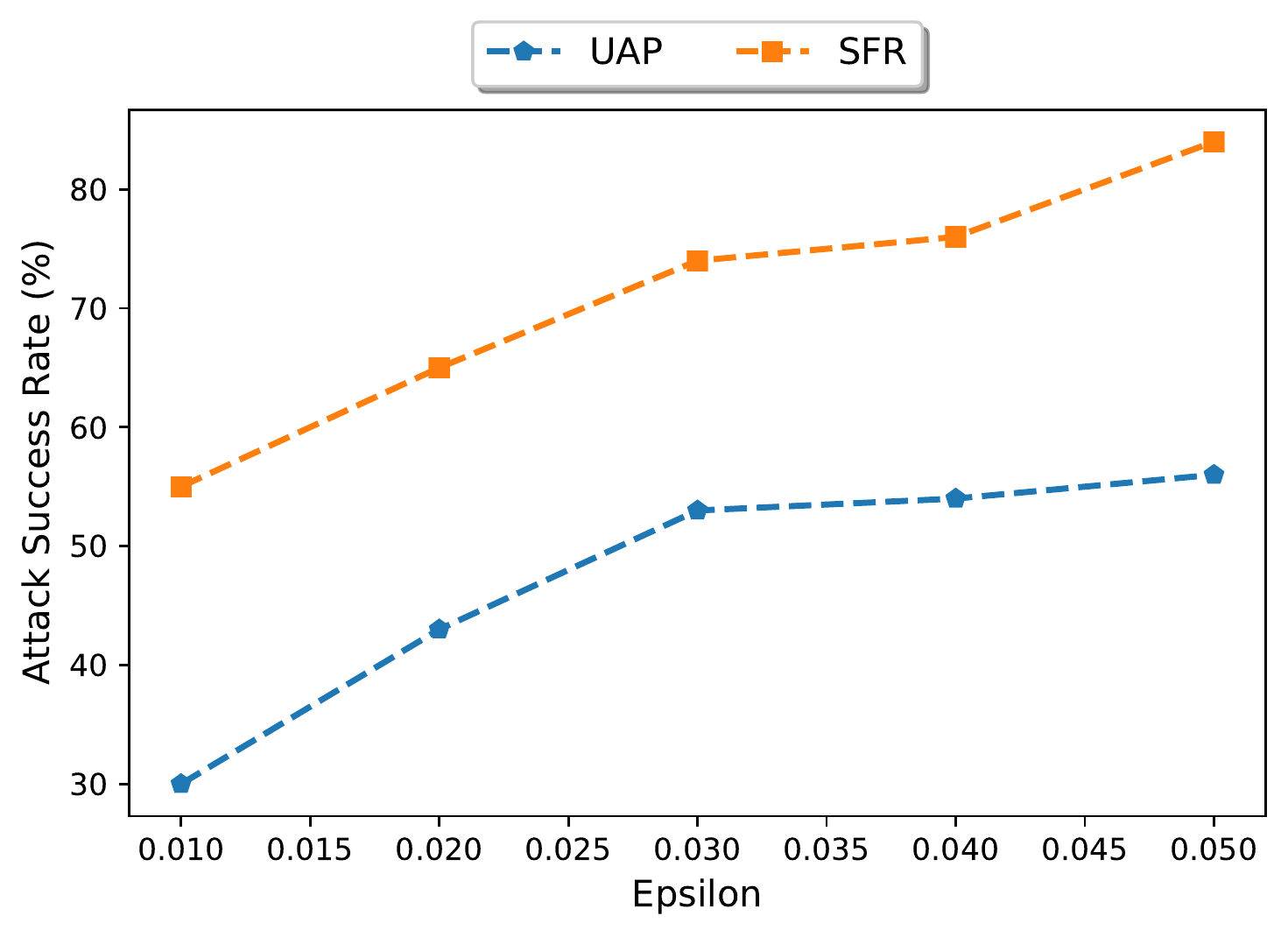}
\caption{Attack success rate of proposed SFR compared to UAP. %\textcolor{red}{is it possible to generate pdf figures instead of png?}
}
\label{FFT}
\end{figure}

Figure \ref{FFT} shows the attack success rate of SFR for different noise budgets comparatively with the baseline UAP.  The results show that SFR can efficiently bypass a defender that uses a filtering stage. For instance, SFR success rate reaches $84\%$  for $0.05$ noise budget while being input-agnostic and incidence delay-resistant, while UAP shows limited efficiency. 

%\begin{table}[h]
%\small \centering
%  \caption{Attack success rate of SFR proposed attack with frequency range restriction.}
%  \label{FFT}
%  \begin{tabular}{ccccc}
%    \toprule
%    \textbf{Epsilon}   & \textbf{0.02} & \textbf{0.03} & \textbf{0.04}& \textbf{0.05}\\
%    \midrule
%        %UAP PGD-based (S)   &    51\%  &    52\% &  55\%   & 58\% \\
%        UAP    &  43\% &  53\% &  54\%  & 56 \%  \\
%        %UAP FGSM-based   &  48\%  &  49\% &   44\%  & 43\% \\
%        %Proposed (alg3) (S) &    73\%  &    78\% &   87\%  &   89\% \\
%        SFR &    65\%  &    74\% &   76\%  &   84\% \\
%  \bottomrule
%\end{tabular}
%\end{table}

%\begin{table}[h]
%\small \centering
%  \caption{Attack success rate of FGSM-based proposed attack with frequency range restriction.}
%  \label{FFT}
%  \begin{tabular}{ccccc}
%    \toprule
%    \textbf{Epsilon}   & \textbf{0.02} & \textbf{0.03} & \textbf{0.04}& \textbf{0.05}\\
%    \midrule
%        UAP FGSM-based   &  48\%  &  49\% &   44\%  & 43\% \\
        %Proposed (alg3) (S) &    \%  &    \% &   \%  &   \% \\
%        Proposed (alg3) &   57\%  &   67\% &  75\%  &  78\% \\
%  \bottomrule
%\end{tabular}
%\end{table}

\subsection{Limits}
While SFR jointly addresses the robustness to incidence time and to filtering defense, its construction opens a new weakness that can be used to detect the adversarial noise. In fact, while generating adversarial noise in the same frequency range of the victim signal bypasses any filter-based pre-processing, it makes the adversarial noise broadcasting detectable by spectrum sensing techniques. In fact, cognitive radars can use different techniques such as energy detection to check for the channel availability. 
While a baseline noise has a wide spectrum and hence a more sparse spectral power distribution, SFR concentrates the spectral components in a relatively narrow band, which makes it detectable by spectrum sensing. 

The next section presents the a-RNA, which keeps all the previous characteristics of SFR while attempting to evade spectrum sensing.

% The limitation of this technique is being detectable by idle spectrum/ 
% spectrum sensing used to have information about its environment and spectrum availability.

\section{Adversarial Radio Noise Attack (a-RNA)}\label{sec:rna}

In this section, we propose a-RNA which is an attack that satisfies the robustness to the previously detailed challenges, i.e., noise incidence delay and filtering, while overcoming SFR limits.

%to reduce the size of the generated patch to avoid detection since noise that is shorter in time is less likely to be observable.

%------------------
\subsection{Approach}
%-----------------
The weakness of SFR is being potentially observable in the spectral domain through idle spectrum sensing. 
The direct reason of this observability consists of the constrained optimization approach that forces the spectral components in a specific range, i.e., the victim operating range. However, the magnitude of the adversarial noise spectral components depend also on the time window for which the noise is broadcasted. For this reason, we propose to add a new constraint on the noise generation in order to limit the magnitude of the frequency components. Specifically, we introduce a time domain window budget that limits the size of the adversarial noise. Therefore, a-RNA attempts to generate an adversarial patch with smaller size rather than using a full size patch to evade detection.
The problem can be formulated as follows:

\begin{equation}\label{eq:arna}\begin{split}
        C(x + shift(\delta, k))) \neq C(x), ~ for ~ most ~ x \sim \mu ~ ~ s.t.\\ 
        \left\Vert \delta \right\Vert_p \leq \epsilon \\
         ~FFT(\delta) \in [f_{min}, f_{max}]\\
         size(\delta) \leq \sigma 
\end{split}
\end{equation}

Where $\epsilon$ controls the magnitude of the noise vector $\delta$ and 
$shift(.)$ is a function expressed by Equation \ref{eq:shift} that quantifies the adversarial patch signal $\delta$, relatively with regard to a target signal $x$ given a time incidence $k \in [0, d] $.
The noise $\delta$ has frequency components constrained in a range defined by $f_{min}$ and $f_{max}$.
The new constraint on $\delta$ is $\sigma$ which represents the budget in terms of noise size in time domain.

Algorithm \ref{attack_alg2} details the noise generation procedure to solve the optimization problem in Equation \ref{eq:arna}. In this algorithm we introduce a temporal mask that extracts a signal that includes $\delta$ with a position $k$ relative to the input signal $x_i$. The signal is padded with trailing zeros to equal the size of the original raw signal. Therefore, the noise is active in a time window of size $s \leq \sigma$ and null elsewhere. 

The function $mask(.)$ could be expressed as:\\

$mask(\delta, k)=\left\{\begin{array}{lll}
\delta(j-k) & \text{if} & j \in[k, k+s] \\
0 & ~ \text{else}
\end{array}\right.$\\

Therefore, $mask()$ function jointly implements the time shift and the size constraint on the noise. This signal is added to the input signal and fed to the $attack()$ function.
We modify $attack()$ to only use the gradient sign of the $[k, k+s]$ signal portion to update the adversarial noise. 

\begin{algorithm}[t]
\caption{Adversarial Radio Noise Attack (a-RNA).}
\label{patch}
\begin{algorithmic}[1]
\State \textbf{Input:} Data points: $X$, classifier: $C$,  noise magnitude: $\epsilon$, number of iterations: $N$, patch size: $s$, raw signal size: $d$.

\State \textbf{Output:} $\delta$ trained patch.\\

\State Initialize $\delta \leftarrow 0$
\While{$iter < N$ }
    \For{each data point $x_i \in X$}
        \If{$C(x_i + mask(\delta, k)) = C(x_i)$}
            \State Select a random location $k$
            \State $x^{adv}_i \leftarrow  attack(C, x_i + mask(\delta, k), y_i, \epsilon, \alpha)$ 
            \If{$C(x_i) != C(x^{adv}_i)$}
                \State $\Delta\delta_i \leftarrow (x^{adv}_i - x_i)[k:k+s]$
            
                \State $\delta \leftarrow \delta + \Delta \delta_i$
                \State $\delta \leftarrow filter(\delta, f_{min}, f_{max} )$
                \State $\delta \leftarrow \mathcal{P}_{l_p, \epsilon}(\delta)$
\EndIf
\EndIf
\EndFor
\State $iter += 1$
\EndWhile

\Function{$mask$}{$\delta$, $k$}
%\State Initialize $Mask \leftarrow 0$
\If{$k <= d - s$}
    \State $Mask = Padding(\delta, (k, d - (k + s)), (0,0))$
\Else
    \State $Mask = Append(zeros(k), \delta [0 : d - k])$
\EndIf

\EndFunction

\end{algorithmic}
\end{algorithm}

\begin{algorithm}[htp]
\caption{$attack$ function. }
\label{attack_alg2}
\begin{algorithmic}[1]
\State \textbf{Input:} a classifier $C$ with loss $J$, noise budget  $\epsilon$, step size $\alpha$, $x$ input image, $y$ label, $m$ number of iterations, $k$ random position, $s$ noise size.
\State \textbf{Output:} $x^{adv}$
%\State Initialize~ $\delta \leftarrow 0$
\State Initialize~ $x^{adv} \leftarrow 0$

\For {i= 0...m-1}
    %\State prediction = $argmax(C(x_{adv})) $
    \State ${x_{i+1}}^{adv} = Clip \{x_i + \alpha sign(\nabla_{x_{i}}^{adv}J_{\theta}(C({x_{i}}^{adv}),y)) \times mask(ones(s),k)\}$
\EndFor

%\State $\delta = x^{adv} - x$
\end{algorithmic}
\end{algorithm}

%---------------------          
\subsection{Results}  
%---------------------
In this section, we evaluate a-RNA from different perspectives:

%---------------------
\noindent\textbf{Impact of patch size on magnitude spectrum. } 
%---------------------
We first investigate the impact of the patch size budget on the magnitude spectrum, which directly implies its detectability using spectrum sensing. The magnitude indicates the strength of the frequency components relative to other components. In Table \ref{magnitude}, we report the average magnitude of the frequency spectrum of each generated patch. We use a-RNA to generate patches with a noise constraint equal to $0.03$. Notice that for a size budget of $600$ (full size), the attack corresponds to SFR case.

Adversarial noises with shorter injection periods have lower magnitudes in the spectral domain, and hence lower chances to be detected. A patch with a size $s = 400$ has a $2\times$ higher magnitude than a patch with a size $s = 100$. A short patch with a size of $50$ samples has around $4\times$ lower average magnitude and $7 \times$ lower maximum magnitude than a SFR patch. \\

\begin{table*}[t!]
\small \centering
  \caption{Magnitude of signals in the spectral domain for different noise sizes (epsilon = 0.03).}
  \label{magnitude}
  \begin{tabular}{ccccccccc}
    \toprule
        \textbf{Patch size} & \textbf{30} & \textbf{50} & \textbf{100} & \textbf{200} & \textbf{300} & \textbf{400} & \textbf{500} & \textbf{600}\\
    \midrule
        Average Magnitude & 0.0555 &  0.0665   & 0.1186  & 0.1368   & 0.1854  & 0.2085    &  0.2343 & 0.2615 \\
        Max Magnitude & 0.4352 &  0.4596   & 0.8011  & 1.1412   & 1.6528  & 2.3231    &  2.8343 & 3.6812\\
  \bottomrule
\end{tabular}
\end{table*}

\begin{figure}[t!]
\centering
\includegraphics[width=\columnwidth]{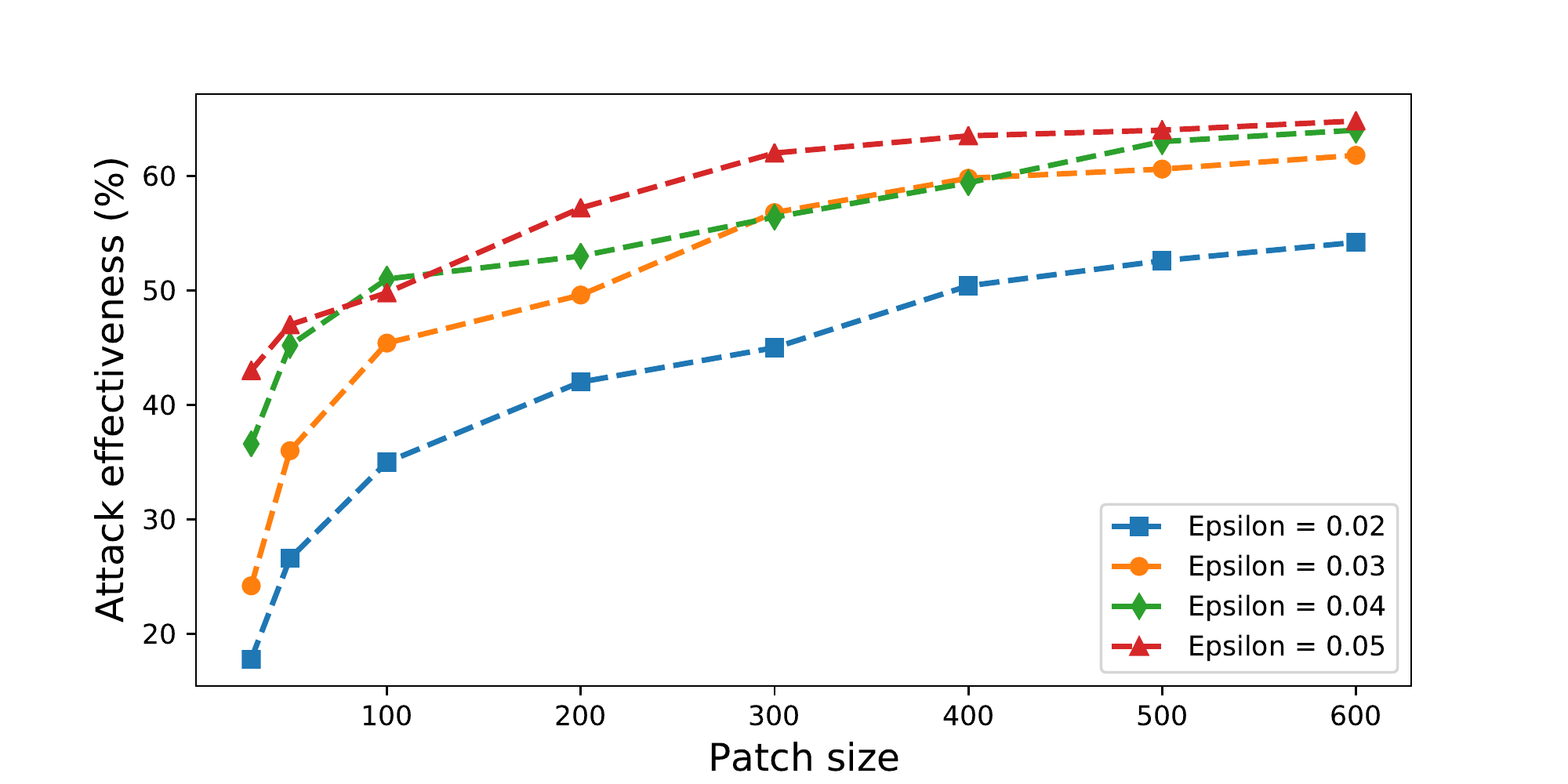}
\caption{Impact of patch size on patch effectiveness under different noise magnitude constraints. }
\label{Fig:patchSize}
\end{figure}

\noindent \textbf{Impact of Patch size on attack efficiency. }
We vary the patch size $s$ from 30 to 600 and investigate its impact on model accuracy. We use algorithm \ref{patch} to train the patches. 
For each test sample, only one patch is diffused at a random time with different noise magnitude ($epsilon$). We notice that the bigger the size of the patch, the more powerful the patch (see Figure \ref{Fig:patchSize}). \\

%\begin{figure*}[t!]
%\centering
%\includegraphics[width=2\columnwidth]{figures/patch_sizevseffectiveness_c.png}
%\caption{Impact of patch size on patch effectiveness under different noise magnitude constraints. }
%\label{Fig:patchSize}
%\end{figure*}

%\begin{figure*}[t!]
%\centering
%\includegraphics[width=2\columnwidth]{figures/repeat_patch_c.png}
%\caption{Continuously broadcast patch effectiveness under different noise magnitude constraints.}
%\label{Fig:contbroa-d}
%\end{figure*}

% \begin{figure}[t!]
% \centering
% \includegraphics[width=\columnwidth]{figures/repeatpatch.pdf}
% \caption{Continuously broadcast patch effectiveness under different noise magnitude constraints.}
% \label{Fig:contbroad}
% \end{figure}

% %---------------------
% \noindent\textbf{Impact of repeating the Patch. }
% %---------------------
% In this experiment, we generate patches with different sizes and under different noise magnitude constraints. The patch is then broadcast continuously. We notice an increase in the success rate of the attack up to $90\%$.
% The patch size is no longer relevant, as illustrated in Figure \ref{Fig:contbroad}. For instance, even a patch with a length $1/22$ of the total raw signal length can achieve $90\%$ success.\\

\begin{figure}[t!]
\centering
\includegraphics[width=\columnwidth]{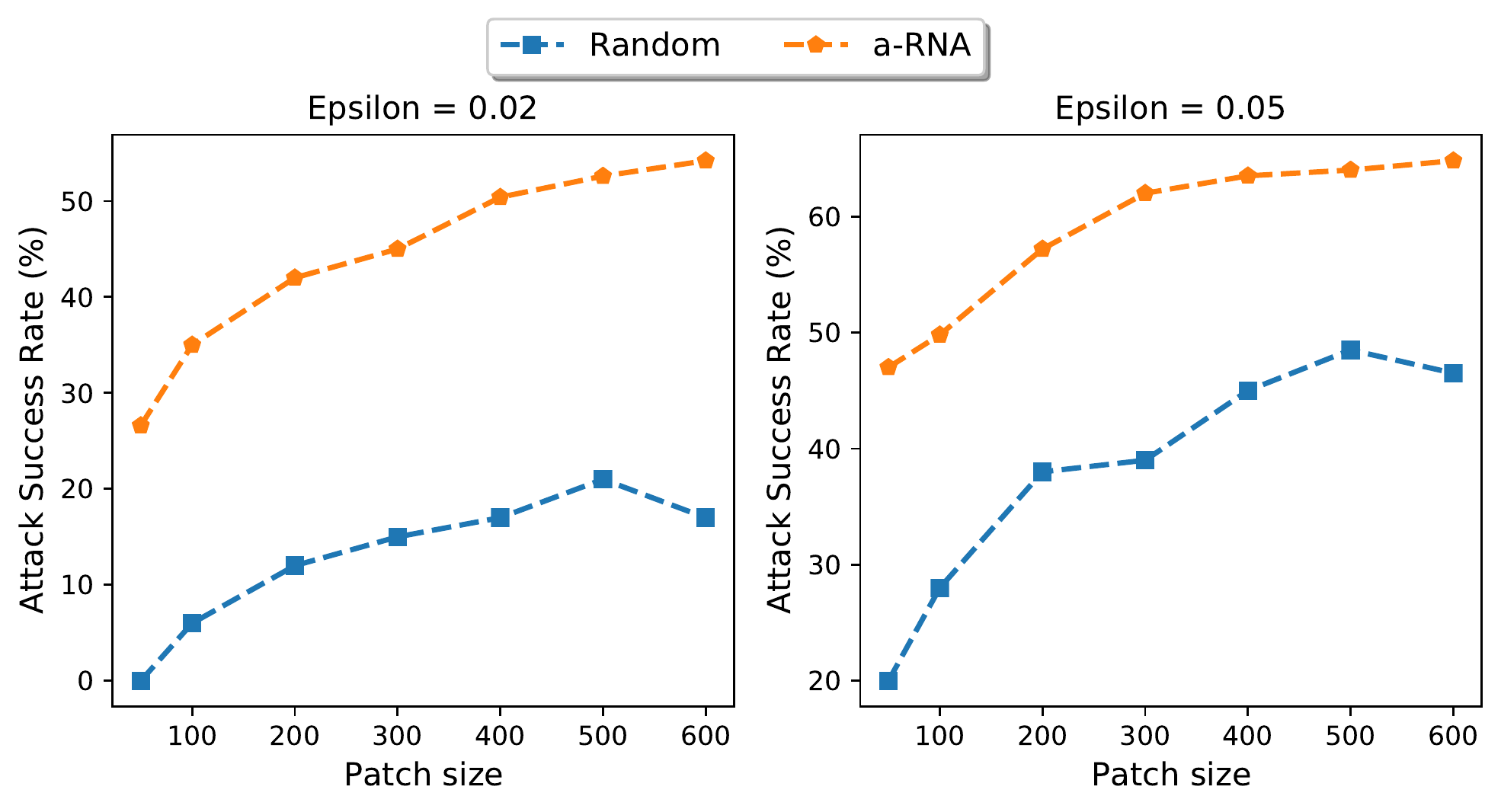}
\caption{Attack success rate of a-RNA compared to random white noise.}
\label{Fig:random}
\end{figure}

%---------------------
\noindent\textbf{Impact of random noise}
%---------------------
In this section we compare our proposed technique a-RNA to injecting random white noise in the wireless channel to cause miss-classification of obstacles.
We use patches of Gaussian white noise with the same magnitude and with different sizes varying from 50 to 600. We use two noise magnitude constraints $0.02$ and $0.05$ for both a-RNA and random white noise. As for the white noise, we use a normal distribution and we vary the mean and the standard deviation to reach the desired noise magnitude.
As shown in Figure \ref{Fig:random}, our technique is more efficient than injecting random noise. For instance, for a patch size equal to 600 and for a noise budget equal to 0.02, a-RNA insures more than $54\%$ attack success rate, however, under the same constraints, the random noise is $17\%$ of the times is successful.

%--------------------------------------------
%\begin{table}[h]
%\small \centering
%  \caption{Attack success rate different injection rates PGD, eps = 0.03.}
%  \label{IR}
%  \begin{tabular}{cccccccc}
%    \toprule
%    \textbf{IR} & \textbf{50} & \textbf{100} & \textbf{200} & \textbf{300} & \textbf{400} & \textbf{500} & \textbf{600}\\
%    \midrule
%                    1    &  34\%    & 45.4\% &  50\%  &  53\% &   56\% & 63\%  & 71\% \\
%                    2    &  40\%    & 53\%   &  56\%  &  83\% &        &       & \\
%                    3    &  45.5\%  & 63\%   &  80\%  &       &        &       & \\  
%                    4    &  50\%    & 68\%   &        &       &        &       & \\
%                    5    &  53.25\% & 78\%   &        &       &        &       & \\  
%                    6    &  54.75\% & 84\%   &        &       &        &       & \\
%                    7    &  56\%    &        &        &       &        &       &\\
%                    8    &  58\%    &        &        &       &        &       &\\
%                    9    &  59\%    &        &        &       &        &       &\\
%                    10   &  62\%    &        &        &       &        &       &\\
%                    11   &  69\%    &        &        &       &        &       &\\
%                    12   &  74\%    &        &        &       &        &       &\\
%                    13   &  87\%    &        &        &       &        &       &\\
%  \bottomrule
%\end{tabular}
%\end{table}

\section{Adversarial Training}\label{sec:def}
%---------------------
%\subsection{Adversarial training}
%---------------------
\begin{figure*}[htp]
\centering
\includegraphics[width=2\columnwidth]{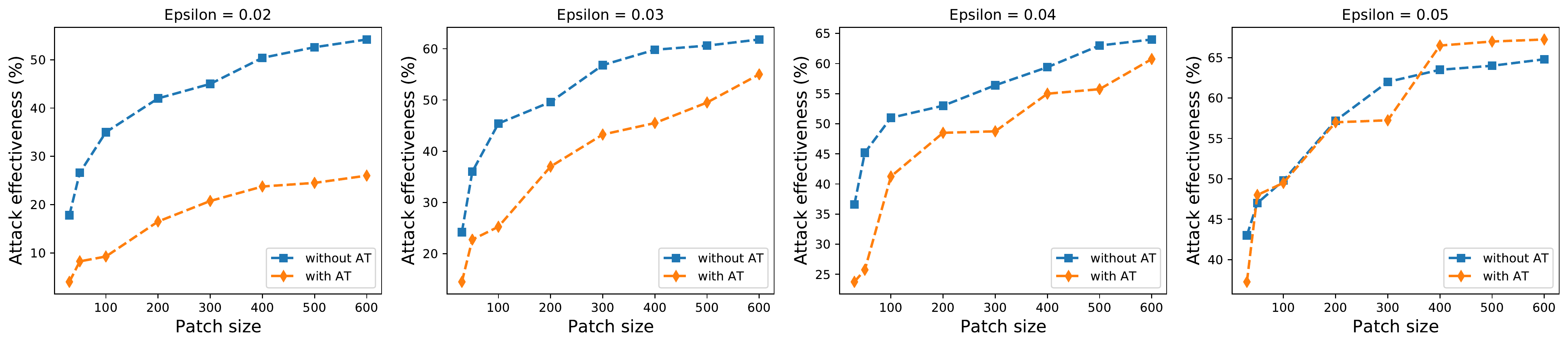}
\caption{Impact of adversarial training (AT) on patch effectiveness.}
\label{Fig:AT}
\end{figure*}

\begin{figure*}[htp]
\centering
\includegraphics[width=2\columnwidth]{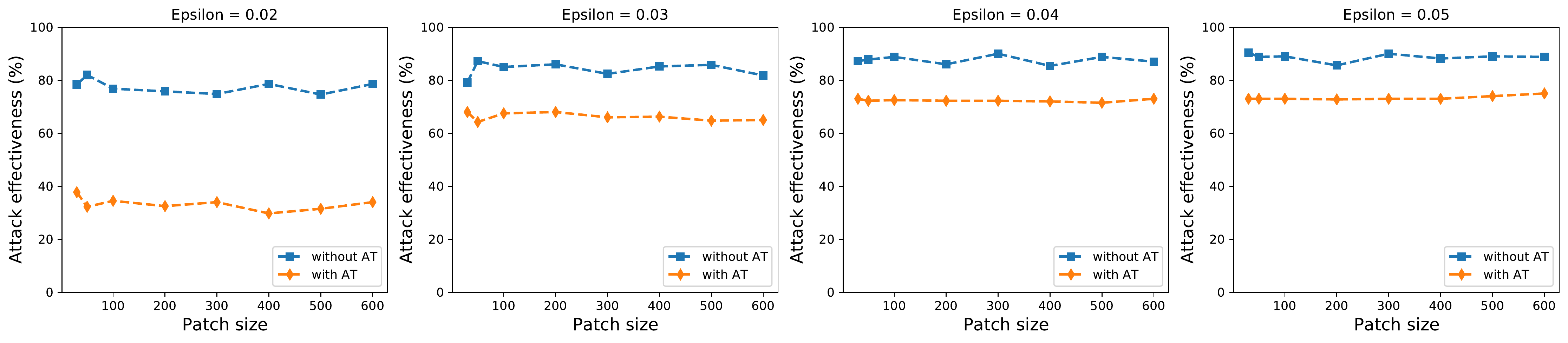}
\caption{Impact of adversarial training (AT) on continuously broadcast patch effectiveness.}
\label{Fig:ATcontbroad}
\end{figure*}

In this section, we evaluate a-RNA in the case of an adversarially trained network.
Adversarial training (AT) \cite{madry2019deep} is a state-of-the-art defense strategy against adversarial attacks. It can be formulated as follows \cite{madry2019deep}: 

\begin{equation}
    \min _{\theta} \mathbb{E}_{(x, y) \sim \mathcal{D}}\left[\max _{\delta \in B(x, \varepsilon)} \mathcal{L}_{c e}(\theta, x+\delta, y)\right]
\end{equation}

Where $\theta$ indicates the parameters of the classifier, $\mathcal{L}_{c e}$ is the cross-entropy loss, $(x, y) \sim \mathcal{D}$ represents the training data sampled from a
distribution $\mathcal{D}$ and $ B(x, \varepsilon)$ is the allowed perturbation
set. The interpretation of this is that the inner maximization problem is finding the
worst-case samples for the given model, and the outer minimization problem is to train a model robust to adversarial
examples \cite{madry2019deep}.

%in which adversarial noises are generated through an adversarial attack, the noises are added to the inputs, and the clean images (without adversarial noises) and the noisy images (added with adversarial noises) are used to train a DNN model.
%A DNN model can learn features of adversarial samples through adversarial training, and by that the classification decision boundary is adjusted so that adding a small amount of adversarial noise makes it difficult to "push" the input samples (images or signals) across the decision boundary.
We used the PGD algorithm \cite{pgd} to solve the inner maximization problem. We set the noise magnitude (epsilon) equal to $0.002$ with a step size of $0.0005$ with a number of iterations equal to $20$. We train the model for $500$ epochs to reach $99\%$ classification accuracy on clean input samples.

In Figure \ref{Fig:AT}, we report the difference in impact of patches generated with a-RNA on adversarially-trained models and undefended models in the case of one random localized patch. As expected, AT decreases the effectiveness of the attack for low adversarial noise magnitude, and is practically bypassed for larger epsilons. Notice that adversarial training for high noise magnitudes may come at a baseline accuracy cost and lead to model utility drop.

To further explore the AT limits, we repeatedly (continuously) broadcast the adversarial patches; results in Figure \ref{Fig:ATcontbroad} show that AT can be evaded with such setting.

In the next section we provide a discussion of our work and point out possible perspectives. 

\section{Discussion}\label{sec:disc}
%discuss the distance Los/NoLos\\
%discuss possible defense ? \\
%discuss higher range sensors 

This paper presents an adversarial attack against UWB-based ML systems for environment perceptions. This is the first work that shows a practical attack against such systems, which are used in several safety-critical applications. We investigate the attack from different perspectives and assumptions. We comprehensively consider real-world constraints such as random incidence time, and different defender models going from basic pre-processing to sophisticated cognitive radio defenders and an adversarially trained model. The proposed approach generates adversarial patches that are: \textbf{(i)} applicable under real-time constraints since they are input-agnostic, \textbf{(ii)} robust to time incidence delay, \textbf{(iii)} robust to filtering techniques, and \textbf{(iv)} can be undetectable even under a defender that deploys both filtering and spectrum sensing. 
Notice that there is a straightforward solution to avoid spectrum sensing: it consists of a cognitive adversarial noise emission by which the attacker broadcasts the adversarial noise exclusively when the victim device is active. This is known in other wireless communication contexts as reactive jamming \cite{reacj1,reacj2,reacj3}. 
In this work, we do not consider this method and focus our effort on the noise generation itself. %Therefore we reduce the observability by adding a constraint on the noise size in the time domain. 

The assumption of this work is a line-of-sight (LoS) communication channel between the adversary and the victim device. While we tested under different distances (with LoS)  and the attack remains efficient as far as the noise magnitude is increased to compensate the path loss, the attack is not successful in no-line-of-sight (No-LoS) settings. In future work, we will study this specific case by integrating the channel model in the noise generation method. 

We believe this work should alert about the potential vulnerabilities of such systems to adversarial attacks in real-world settings, especially for critical applications. We hope these results would encourage the community to further investigate this direction.

\section{ Related Work}\label{sec:related}
%UWB-based systems that rely on deep learning techniques were mainly employed for indoor applications such as: activity recognition, indoor people localization and recognition of movements during sleep. Few are those who have considered it for outdoor applications.. 
%2D radar imaging based systems are no longer considered a signal processing problem, but rather an image processing problem. Otherwise, 1D UWB signals are likewise employed with deep learning for indoor applications.

Computer vision is the mainstream application that caught the attention of the community from adversarial machine learning perspective \cite{CW,fgsm,pgd}. However, the vulnerability of ML in other application domains has also been explored. For example, several papers have explored adversarial attacks on Lidar sensors \cite{lidarCVPR20} or electrocardiograms \cite{ecg_nature}. Some of the literature on automatic speech/speaker recognition have been designed under practical real-world assumptions \cite{asr_ufl, patch,multi,face,ASR}. Yet, most of the works require very specific settings to be applicable.  
%yet most of the literature 
%works are practical in real life. automatic speech/speaker recognition \cite{ASR,bob,asr_ufl}, . While some of the on speech  work been designed under practical real-world conditions \cite{asr_ufl, patch,multi,face,ASR}, they are mainly focusing . 
%, most of the works require very specific settings to be applicable.  

Several application of wireless communication use ML. These radio applications represent a potential target for adversarial attacks that are practical under real-world conditions. Recent work proposed crafting adversarial attacks against ML-based wireless systems. However, most of the
existing works \cite{CCS21_1, CCS21_14,CCS21_15,CCS21_32,CCS21_33} use a input-specific perturbation to attack the target wireless model and do
not provide the key properties for a practical attack, i.e., robustness against realistic conditions and active defenses. A recent paper present an attack against DNN-based wireless communication system \cite{CCS21}. This paper is the closest to our work and considers three applications: autoencoder communication, modulation recognition and channel estimation. While the authors presented an input-agnostic noise and discuss its robustness to potential defense techniques, the application case prevents the attack from being implementable in real life, and this is mainly due to the complex propagation channel of the adversarial noise itself.

Among wireless applications, radars are used for environment perception, and similar to other ML-based applications, they are vulnerable to adversarial examples. More specifically, short range systems such as some radar technologies represent a highly practical attack setting because of the possibility of line-of-sight transmission conditions. Few papers in the literature target these systems \cite{xband,fmcw}. Authors in \cite{xband} consider X-band spotlight mode radar which is utilized for hundreds of kilometers range and not practical for injecting adversarial noise due to the channel complexity. \cite{fmcw} targets short-range Frequency-modulated continuous-wave (FMCW) radars and is the closest paper to our work. However, FMCW radars give only velocity and range information, while UWB delivers a complete signature of the obstacle. 

In this work, we propose an input-agnostic, undetectable, and robust adversarial attack against ML-based UWB radar systems. We design tailored universal patches to perform the attack and discuss their efficiency from practical perspectives. The short range aspect of the UWB radars represent a practical case due to the simple channel model.

%X-band spotlight mode radar: \cite{xband}\\

%Frequency-modulated continuous-wave (FMCW): \cite{fmcw}

\section{Conclusion}
We present a new adversarial radio noise attack (a-RNA) on UWB radars to generate a noise robust by design against realistic conditions and adaptive against defensive countermeasures. To our knowledge, this is the first approach to generate such practical attacks against DNN-based UWB systems. We believe a-RNA should alert the community about the feasibility of real-world attacks against radar systems.  

%leverages unique characteristics of Ultra-Wide Band radar signals to inject adversarial noise in a radar-based environment perception system.
%-------------------------------------------------------------------------------
\bibliographystyle{plain}
\bibliography{bib}

\begin{thebibliography}{10}

\bibitem{asr_ufl}
Hadi Abdullah, Washington Garcia, Christian Peeters, Patrick Traynor, Kevin
  Butler, and Joseph Wilson.
\newblock {Practical Hidden Voice Attacks against Speech and Speaker
  Recognition Systems}.
\newblock In {\em Network and Distributed System Security Symposium (NDSS)},
  2019.

\bibitem{CCS21_1}
Abdullatif Albaseer, Bekir~Sait Ciftler, and Mohamed~M. Abdallah.
\newblock Performance evaluation of physical attacks against e2e autoencoder
  over rayleigh fading channel.
\newblock In {\em 2020 IEEE International Conference on Informatics, IoT, and
  Enabling Technologies (ICIoT)}, pages 177--182, 2020.

\bibitem{CW}
Anish Athalye, Nicholas Carlini, and David Wagner.
\newblock Obfuscated gradients give a false sense of security: Circumventing
  defenses to adversarial examples.
\newblock In Jennifer Dy and Andreas Krause, editors, {\em Proceedings of the
  35th International Conference on Machine Learning}, volume~80 of {\em
  Proceedings of Machine Learning Research}, pages 274--283. PMLR, 10--15 Jul
  2018.

\bibitem{CCS21}
Alireza Bahramali, Milad Nasr, Amir Houmansadr, Dennis Goeckel, and Don
  Towsley.
\newblock Robust adversarial attacks against dnn-based wireless communication
  systems.
\newblock In {\em Proceedings of the 2021 ACM SIGSAC Conference on Computer and
  Communications Security}, CCS '21, page 126–140, New York, NY, USA, 2021.
  Association for Computing Machinery.

\bibitem{bob}
Guangke Chen, Sen Chenb, Lingling Fan, Xiaoning Du, Zhe Zhao, Fu~Song, and Yang
  Liu.
\newblock Who is real bob? adversarial attacks on speaker recognition systems.
\newblock In {\em 2021 IEEE Symposium on Security and Privacy (SP)}, pages
  694--711, 2021.

\bibitem{fgsm}
Ian~J. Goodfellow, Jonathon Shlens, and Christian Szegedy.
\newblock Explaining and harnessing adversarial examples, 2014.

\bibitem{guesmi2021sit}
Amira Guesmi, Ihsen Alouani, Mouna Baklouti, Tarek Frikha, and Mohamed Abid.
\newblock Sit: Stochastic input transformation to defend against adversarial
  attacks on deep neural networks.
\newblock {\em IEEE Design \& Test}, 2021.

\bibitem{defensiveapproximation}
Amira Guesmi, Ihsen Alouani, Khaled~N. Khasawneh, Mouna Baklouti, Tarek Frikha,
  Mohamed Abid, and Nael Abu-Ghazaleh.
\newblock {\em Defensive Approximation: Securing CNNs Using Approximate
  Computing}, page 990–1003.
\newblock Association for Computing Machinery, New York, NY, USA, 2021.

\bibitem{ecg_nature}
Xintian Han, Yuxuan Hu, Luca Foschini, Larry Chinitz, Lior Jankelson, and
  Rajesh Ranganath.
\newblock Deep learning models for electrocardiograms are susceptible to
  adversarial attack.
\newblock {\em Nature Medicine}, 26(3):360--363, 2020.

\bibitem{xband}
Teng Huang, Yongfeng Chen, Bingjian Yao, Bifen Yang, Xianmin Wang, and Ya~Li.
\newblock Adversarial attacks on deep-learning-based radar range profile target
  recognition.
\newblock {\em Information Sciences}, 531:159--176, 2020.

\bibitem{com_lett}
Changhui Jiang, Jichun Shen, Shuai Chen, Yuwei Chen, Di~Liu, and Yuming Bo.
\newblock Uwb nlos/los classification using deep learning method.
\newblock {\em IEEE Communications Letters}, 24(10):2226--2230, 2020.

\bibitem{CCS21_15}
Brian Kim, Yalin~E. Sagduyu, Kemal Davaslioglu, Tugba Erpek, and Sennur Ulukus.
\newblock Over-the-air adversarial attacks on deep learning based modulation
  classifier over wireless channels.
\newblock In {\em 2020 54th Annual Conference on Information Sciences and
  Systems (CISS)}, pages 1--6, 2020.

\bibitem{CCS21_14}
Brian Kim, Yalin~E. Sagduyu, Kemal Davaslioglu, Tugba Erpek, and Sennur Ulukus.
\newblock Channel-aware adversarial attacks against deep learning-based
  wireless signal classifiers, 2021.

\bibitem{face}
Stepan Komkov and Aleksandr Petiushko.
\newblock Advhat: Real-world adversarial attack on arcface face id system.
\newblock In {\em 2020 25th International Conference on Pattern Recognition
  (ICPR)}, pages 819--826, 2021.

\bibitem{its_review}
Yongqiang Lu, Hongjie Ma, Edward Smart, and Hui Yu.
\newblock Real-time performance-focused localization techniques for autonomous
  vehicle: A review.
\newblock {\em IEEE Transactions on Intelligent Transportation Systems}, pages
  1--19, 2021.

\bibitem{pgd}
Aleksander Madry, Aleksandar Makelov, Ludwig Schmidt, Dimitris Tsipras, and
  Adrian Vladu.
\newblock Towards deep learning models resistant to adversarial attacks, 2017.

\bibitem{madry2019deep}
Aleksander Madry, Aleksandar Makelov, Ludwig Schmidt, Dimitris Tsipras, and
  Adrian Vladu.
\newblock Towards deep learning models resistant to adversarial attacks, 2019.

\bibitem{exp}
Julien Maitre, Kévin Bouchard, Camille Bertuglia, and Sébastien Gaboury.
\newblock Recognizing activities of daily living from uwb radars and deep
  learning.
\newblock {\em Expert Systems with Applications}, 164:113994, 2021.

\bibitem{olimp}
Amira Mimouna, Ihsen Alouani, Anouar Ben~Khalifa, Yassin El~Hillali, Abdelmalik
  Taleb-Ahmed, Atika Menhaj, Abdeldjalil Ouahabi, and Najoua~Essoukri
  Ben~Amara.
\newblock Olimp: A heterogeneous multimodal dataset for advanced environment
  perception.
\newblock {\em Electronics}, 9(4), 2020.

\bibitem{entropy}
Amira Mimouna, Anouar~Ben Khalifa, Ihsen Alouani, Najoua Essoukri~Ben Amara,
  Atika Rivenq, and Abdelmalik Taleb-Ahmed.
\newblock Entropy-based ultra-wide band radar signals segmentation for multi
  obstacle detection.
\newblock {\em IEEE Sensors Journal}, 21(6):8142--8149, 2021.

\bibitem{Universal}
S.~{Moosavi-Dezfooli}, A.~{Fawzi}, O.~{Fawzi}, and P.~{Frossard}.
\newblock Universal adversarial perturbations.
\newblock In {\em 2017 IEEE Conference on Computer Vision and Pattern
  Recognition (CVPR)}, pages 86--94, 2017.

\bibitem{reacj3}
Danh Nguyen, Cem Sahin, Boris Shishkin, Nagarajan Kandasamy, and Kapil~R.
  Dandekar.
\newblock A real-time and protocol-aware reactive jamming framework built on
  software-defined radios.
\newblock In {\em Proceedings of the 2014 ACM Workshop on Software Radio
  Implementation Forum}, SRIF '14, page 15–22, New York, NY, USA, 2014.
  Association for Computing Machinery.

\bibitem{fmcw}
Utku Ozbulak, Baptist Vandersmissen, Azarakhsh Jalalvand, Ivo Couckuyt, Arnout
  {Van Messem}, and Wesley {De Neve}.
\newblock Investigating the significance of adversarial attacks and their
  relation to interpretability for radar-based human activity recognition
  systems.
\newblock {\em Computer Vision and Image Understanding}, 202:103111, 2021.

\bibitem{reacj2}
Konstantinos Pelechrinis, Marios Iliofotou, and Srikanth~V. Krishnamurthy.
\newblock Denial of service attacks in wireless networks: The case of jammers.
\newblock {\em IEEE Communications Surveys Tutorials}, 13(2):245--257, 2011.

\bibitem{ASR}
Yao Qin, Nicholas Carlini, Garrison Cottrell, Ian Goodfellow, and Colin Raffel.
\newblock Imperceptible, robust, and targeted adversarial examples for
  automatic speech recognition.
\newblock In Kamalika Chaudhuri and Ruslan Salakhutdinov, editors, {\em
  Proceedings of the 36th International Conference on Machine Learning},
  volume~97 of {\em Proceedings of Machine Learning Research}, pages
  5231--5240. PMLR, 09--15 Jun 2019.

\bibitem{CCS21_32}
Meysam Sadeghi and Erik~G. Larsson.
\newblock Adversarial attacks on deep-learning based radio signal
  classification.
\newblock {\em IEEE Wireless Communications Letters}, 8(1):213--216, 2019.

\bibitem{CCS21_33}
Meysam Sadeghi and Erik~G. Larsson.
\newblock Physical adversarial attacks against end-to-end autoencoder
  communication systems.
\newblock {\em IEEE Communications Letters}, 23(5):847--850, 2019.

\bibitem{Saito2003}
Akihiko Saito, Hiroshi Harada, and Atsuhiro Nishikata.
\newblock Development of band pass filter for ultra wideband (uwb)
  communication systems.
\newblock {\em IEEE Conference on Ultra Wideband Systems and Technologies,
  2003}, pages 76--80, 2003.

\bibitem{reacj1}
Mario Strasser, Boris Danev, and Srdjan \v{C}apkun.
\newblock Detection of reactive jamming in sensor networks.
\newblock {\em ACM Trans. Sen. Netw.}, 7(2), sep 2010.

\bibitem{multi}
Bilel Tarchoun, Ihsen Alouani, Anouar Ben~Khalifa, and Mohamed~Ali Mahjoub.
\newblock Adversarial attacks in a multi-view setting: An empirical study of
  the adversarial patches inter-view transferability.
\newblock In {\em 2021 International Conference on Cyberworlds (CW)}, pages
  299--302, 2021.

\bibitem{patch}
Simen Thys, Wiebe Van~Ranst, and Toon Goedeme.
\newblock Fooling automated surveillance cameras: Adversarial patches to attack
  person detection.
\newblock In {\em Proceedings of the IEEE/CVF Conference on Computer Vision and
  Pattern Recognition (CVPR) Workshops}, June 2019.

\bibitem{lidarCVPR20}
James Tu, Mengye Ren, Sivabalan Manivasagam, Ming Liang, Bin Yang, Richard Du,
  Frank Cheng, and Raquel Urtasun.
\newblock Physically realizable adversarial examples for lidar object
  detection.
\newblock In {\em Proceedings of the IEEE/CVF Conference on Computer Vision and
  Pattern Recognition (CVPR)}, June 2020.

\bibitem{umain}
{UMAIN I}nc.
\newblock {HST-D3} evaluation kit.
\newblock \url{https://umain.en.ec21.com/}.

\bibitem{wang2022}
Changqiang Wang, Aigong Xu, Xin Sui, Yushi Hao, Zhengxu Shi, and Zhijian Chen.
\newblock A seamless navigation system and applications for autonomous vehicles
  using a tightly coupled gnss/uwb/ins/map integration scheme.
\newblock {\em Remote Sensing}, 14(1):27, 2022.

\bibitem{pbform}
Xiaoyong Yuan, Pan He, Qile Zhu, Rajendra~Rana Bhat, and Xiaolin Li.
\newblock Adversarial examples: Attacks and defenses for deep learning.
\newblock {\em CoRR}, abs/1712.07107, 2017.

\bibitem{Zetik}
Rudolf Zetik, Jurgen Sachs, and Reiner~S. Thoma.
\newblock Uwb short-range radar sensing - the architecture of a baseband,
  pseudo-noise uwb radar sensor.
\newblock {\em IEEE Instrumentation Measurement Magazine}, 10(2):39--45, 2007.

\end{thebibliography}

% \newpage
% \section{Appendix}
% %---------------------
% \noindent\textbf{Impact of random noise}
% %---------------------
% In this section we compare our proposed technique a-RNA to injecting random white noise in the wireless channel to cause miss-classification of obstacles.
% We use patches of Gaussian white noise with the same magnitude and with different sizes. We use two noise magnitude constraints $0.02$ and $0.05$. As shown in Figure \ref{Fig:random}, our technique is more efficient than injecting random noise. 
% \begin{figure}
% \centering
% \includegraphics[width=\columnwidth]{figures/random.pdf}
% \caption{Attack success rate of SFR compared to random white noise.%\textcolor{red}{SFR? ou a-RNA?}}
% \label{Fig:random}
% \end{figure}

%%%%%%%%%%%%%%%%%%%%%%%%%%%%%%%%%%%%%%%%%%%%%%%%%%%%%%%%%%%%%%%%%%%%%%%%%%%%%%%%
\end{document}